%% file: rainy.tex
\documentclass{jfm} 
\pdfoutput=1
\usepackage{xspace}
\usepackage{graphicx}
\usepackage{natbib}
\usepackage{ifpdf,ifxetex,ifluatex} 
\usepackage{amsmath}
\usepackage{subcaption}
\usepackage[full]{textcomp}
\usepackage{verbatim}
\usepackage[scaled=1.05]{newtxtext}
\usepackage[scaled=1.05, varg]{newtxmath}
\usepackage{microtype}
\usepackage{bm}
\usepackage{hyperref} \hypersetup{pdfencoding=auto,unicode, psdextra, hidelinks}

\renewcommand{\boldsymbol}{\bm}
\newcommand{\figref}[1]{Fig.~\ref{#1}}

\nonstopmode

\newcommand{\eqtag}[1]%
    {\renewcommand{\theequation}{\arabic{section}.\arabic{equation}#1}}

\renewcommand{\H}{\mathcal H}

\usepackage{color}
\definecolor{light-gray}{gray}{0.5}
\definecolor{blue}{rgb}{0.0,0.0,1.0}
\definecolor{green}{rgb}{0.0,0.5,0.0}
\definecolor{red}{rgb}{1.0,0.0,0.0}
\definecolor{cyan}{rgb}{0.0,0.75,0.75}
\definecolor{magenta}{rgb}{0.75,0.0,0.75}
\definecolor{yellow}{rgb}{0.75,0.75,0.0}

\newcommand{\avg}[1]{\langle{#1}\rangle}

\newcommand{\D}{\Delta}

\newcommand{\Ra}{\mathit{Ra}}

\newcommand{\RB}{Rayleigh--B\'enard\xspace}
\newcommand{\rb}{rainy-B\'enard\xspace}

\newcommand{\secref}[1]{\S{\kern 1pt}\ref{#1}}

\title[Rainy--B\'enard Convection]{A Simple System for Moist Convection: \\ The Rainy-B\'enard Model}

\author[]%
{Geoffrey K. Vallis,$^1$
Douglas J. Parker$^2$ and Steven M. Tobias$^3$}

\affiliation{$^1$Department of Mathematics, University of Exeter, Exeter, EX4 4QF, UK \\
$^2$School of Earth and Environment, University of Leeds,  Leeds LS2 9JT, UK \\
$^3$Department of Applied Mathematics, University of Leeds, Leeds LS2 9JT, UK}

\pubyear{2019}
\volume{111}
\pagerange{1--26}
\input{mypreamble}
\begin{document} 

\maketitle

\begin{abstract}
 \RB convection is one of the most well-studied models in fluid mechanics. Atmospheric convection, one of the most important components of the climate system, is by comparison complicated and poorly understood. A key attribute of atmospheric convection is the  buoyancy source provided by the condensation of water vapour, but the presence of radiation, compressibility, liquid water and ice further complicate the system and our understanding of it.  In this paper we present an idealized model of moist convection by taking the Boussinesq limit of the ideal gas equations and adding a condensate that obeys a simplified Clausius--Clapeyron relation. The system allows moist convection to be explored at a fundamental level and reduces to the classical \RB model if the latent heat of condensation is taken to be zero. The model has an exact, Rayleigh-number independent `drizzle' solution in which the diffusion of water vapour from a saturated lower surface is balanced by condensation, with the temperature field (and so the saturation value of the moisture) determined self-consistently by the heat released in the condensation.   This state is the moist analogue of the conductive solution in the classical problem. We numerically determine the linear stability properties of this solution as a function of Rayleigh number and a nondimensional latent-heat parameter. We also present some two-dimensional, time-dependent, nonlinear solutions at various values of Rayleigh number and the nondimensional condensational  parameters. At sufficiently low Rayleigh number the system converges to the drizzle solution, and we find no evidence that two-dimensional self-sustained convection can occur when that solution is stable.  The flow transitions from steady to turbulent as the Rayleigh number or the effects of condensation are increased, with plumes triggered by gravity waves emanating from other plumes. The interior dries as the level of turbulence increases, because the plumes entrain more dry air and because the saturated boundary layer at the top becomes thinner. The flow develops a broad relative humidity minimum in the domain interior, only weakly dependent on Rayleigh number when that is high.  
\end{abstract}


\nocite{Vallis17}
\section{Introduction} \label{sec:intro} 
Convection is ubiquitous in fluids,  certainly in geophysical and astrophysical settings. In Earth's atmosphere convection is responsible for the towering cumulonimbus of the tropics and plays a major role in mid-latitude storm systems \citep{Emanuel94, Smith13}, and in the ocean convection ventilates the deep abyss \citep[e.g.,][]{Marshall_Schott99}. Convection in the Earth's mantle leads to continental drift \citep{Parsons_MacKenzie78}, and deeper in the fluid outer core convective dynamos produce the magnetic field of Earth and other planets  \citep{Schubert_Soderlund11}. In stars convection enables the heat released by fusion to escape the burning core maintaining a quasi-steady state and leads to stellar magnetic field generation \citep{Brun_Browning17}.   

A classic problem is that of \RB convection. Here, in the usual configuration, a Boussinesq fluid is confined between two horizontal plates held at different, constant (in both space and time) temperatures, with the fluid having uniform diffusivities and viscosities, and literally hundreds of papers have been written on this or closely connected problems --- see \citet{Ahlers_etal09} or \citet{Chilla_Schumacher12} for reviews.  A solution of no motion then exists, with a uniform gradient of temperature between bottom and top, and temperature diffusing from one plate to the other.  If the lower plate is at a higher temperature than the upper one, and the Rayleigh number is sufficiently high, then the diffusive state is linearly unstable \citep{Rayleigh16, Chandrasekhar61, Drazin_Reid81} and convection results. Depending on the value of the controlling nondimensional parameters (the Rayleigh number, the Prandtl number and the aspect ratio) the ensuing motion may be steady or turbulent, or somewhere in between, with the level of turbulence normally increasing as the Rayleigh number increases. The convecting flow may also coalesce and/or aggregate into various patterns, many of which have been extensively studied numerically and experimentally. 

Various theories have also been proposed for how the heat transport scales with the Rayleigh number. In particular, a so-called ultimate regime has been proposed in which the heat transport becomes independent of the diffusivity and viscosity \citep{Kraichnan62, Spiegel71}, much as the energy dissipation in three-dimensional turbulence becomes independent of viscosity at high Reynolds numbers. However, evidence that an ultimate regime exists in the standard problem is mixed, even at very high Rayleigh number \citep{Ahlers_etal09} and in some configurations the ultimate regime provably cannot exist \citep{Whitehead_Doering11}. One is left with the somewhat uncomfortable notion that the heat transport in \RB convection may always depend on molecular properties, no matter how turbulent the flow. The fundamental difference in this regard between \RB convection and homogeneous turbulence stems from the fact that the buoyancy must enter the fluid through a thin boundary layer in which diffusivity is unavoidably important. If in numerical experiments one eliminates the boundary layer by making the flow periodic in the vertical  \citep{Lohse_Toschi03}, or if in laboratory experiments one breaks the boundary layer by making the surface rough  \citep{Roche_etal01}, then an ultimate state for those systems may emerge.

As fundamental and important as these matters are, progress in the theory of atmospheric convection has, with a few exceptions, proceeded largely independently of them.  One reason for this is that Earth's atmosphere contains a condensate, water.  A wet lower surface provides a moisture source, and when the water vapour condenses heat is released, enhancing the convection. The consequences of this are considerable, with the most apparent being that the critical lapse rate for convection is no longer simply that corresponding to a vertical buoyancy (or potential temperature) gradient of zero (neglecting viscous and diffusive effects); rather, a profile that is stable when dry may be unstable when saturated \citep[e.g.,][]{Emanuel94, Ambaum10}.  The `saturated adiabatic lapse rate' gives this critical lapse rate, which in Earth's atmosphere varies between about 3\dg/km and 9.8\dg/km, whereas the `dry adiabatic lapse rate' for an ideal gas is given by $g/\cp$ and is 9.8\dg/km on Earth corresponding to zero vertical gradient of potential temperature (which is a proxy for buoyancy).  Moisture also brings with it the possibility of \textit{conditional instability.}  Here, a fluid parcel containing moisture may be stable to small perturbations, but if the parcel is lifted and the water vapour condenses, the release of latent heat may then be sufficient to destabilize the fluid.   Furthermore,  after condensation the water droplets may stay suspended as a cloud (affecting the radiation budget), or freeze to form ice, or the moisture may fall as rain.  The rain drops may partially re-evaporate before reaching the ground, leading to `cold pools' --  regions of evaporatively cooled air that reach the surface and then spread out as density currents \citep[e.g.,][]{Tompkins01}, and which are sometimes regarded as key ingredients in convective organization.  A great many types of behaviour have been seen -- including clustering and aggregation and other patterns. Moisture is not essential to organized behaviour,  since various patterns occur in dry \RB convection \citep{Golubitsky_etal84, Cross_Hohenberg93}, but is likely important in determining  the nature of the organization. 

Although the presence of condensation is the main difference between \RB convection and atmospheric convection, it is not the only one. \RB convection is, by tradition, dry,  uses a Boussinesq fluid between two plates, and diffusion  of heat from the boundaries provides the only source and sink of buoyancy. As well as being (frequently) moist, atmospheric convection occurs in an ideal gas, has no upper boundary and radiation provides another source of buoyancy --- both heating the fluid  near the ground and, perhaps more significantly, cooling the fluid internally.  The radiative cooling  brings an up-down asymmetry \citep{Parodi_etal03, Berlengiero_etal12} that may be as significant as moisture itself in so doing, and may be at least partially responsible for removing the bottleneck of molecular diffusivity in the boundary.  Shear induced turbulence at the lower boundary is also likely important in this regard.   These various complications have led to theories of atmospheric convection becoming separated from the theories of \RB convection, with respect to both boundary layers and deep convection.  For example, the similarity theory of \citet{Monin_Obukhov54} has provided a very well-verified foundation for parameterizations of the atmospheric boundary layer in some conditions, and it makes no reference to molecular values at all.  It does, however, make reference to a roughness length and a `Monin--Obukhov' length to take into account buoyancy effects.  This remarkably useful theory in the field has found little direct application in laboratory studies in which the free convective limit is the main interest,  but it does point to the importance of surface roughness, and indeed laboratory experiments almost always find that the Nusselt number increases with roughness \citep{Chilla_Schumacher12}.   Thus, although the two fields may be disjoint, they are not in contradiction.

Away from the boundary layer, theories of atmospheric convection have tended to draw more on plume theories and (more global in nature) quasi-equilibrium ideas \citep[e.g.,][]{Emanuel_etal94} that have their origins in \citet{Scorer_Ludlam53}, \citet{Morton_etal56} and \citet{Ludlam66}, whereas theories of \RB turbulence tend to be more statistically based and/or seek scaling relations (Kraichnan, op cit, \citealt{Grossmann_Lohse00}).  The numerical models used in atmospheric convection also vary widely, with different choices of microphysics, radiation schemes and resolution --- with the resolution being far from that needed to use molecular values of viscosity and diffusion, or even from resolving the cloud plumes \citep[e.g.,][and true today]{Bryan_etal03}. Various eddy viscosity and subgrid parameterization schemes are perforce used  for both the  free atmosphere and the boundary layer (e.g., \citet{Smagorinsky63}, \citet{Mellor_Yamada74} and $k$--$\epsilon$ models) and depending on the resolution and choices used the models may be labelled large-eddy models or cloud resolving models. The parameters that are known to be important in \RB convection, specifically the Rayleigh number and the Prandtl number, then play no direct role in most theories (or numerical models) of atmospheric convection, and the differences between the various models often makes it difficult to know whether the behaviour seen in a particular case (e.g., convective aggregation) but not in another is a true effect or an artifact of a particular model.  The consequence is that reproducibility is difficult to achieve unless the numerical code itself is duplicated, and model intercomparison projects are used to compare models.  In some contrast, the transition to turbulence and indeed some of the turbulent properties themselves of dry \RB convection are now fairly well modelled. 

However, dry \RB convection is of itself a poor model of atmospheric convection, because the simplifications of the model become oversimplifications in an atmospheric context.  It is one of the purposes of this paper to help bridge the gap between a simple but not directly relevant model (the \RB one) and complex cloud resolving models with less well understood behaviour. Along these lines, Emanuel (1994) begins his book on atmospheric convection with a discussion of plumes and the \RB problem, and \citet{Bretherton87, Bretherton88}, \citet{Spyksma_etal06}, and \citet{Pauluis_Schumacher10} and \citet{Schumacher_Pauluis10} all make links between the two systems. Bretherton was interested in the properties of non-precipitating convection and included liquid water in a fluid contained between two plates. Spyksma \textit{et al.} performed  bubble simulations with a similar model,  implemented in a triply-periodic domain. Pauluis and Schumacher also kept liquid water and used a piecewise linear equation of state in which the buoyancy derivatives depend only on whether a parcel is saturated or not.  Precipitating convection was the object of the idealized study by \citet{Hernandez-etal13}, who presented a model in which water vapour converted to rain water that fell and potentially re-evaporated.  A number of studies have also examined the problem of two-phase \RB convection, for example when a fluid boils \citep{Schmidt_etal11, Lakkaraju_etal13} or in mantle convection \citep{Christensen95}.  Note that no system enables one to model a turbulent convection problem in a realistic geophysical or astrophysical setting using molecular values for diffusion and viscosity. 
 
In this paper our goals are to present a model that represents an essence of moist convection, in order that the system may be explored at a fundamental level, and to begin to study that model.  We do this by adding a single, active, condensate to the \RB model; all the condensate is removed immediately on saturation (there is no liquid phase), so providing a buoyancy source, and diffusion from the lower boundary replenishes the humidity. This `\rb' model is thus relatively simple (in construct, if not behaviour) and is wholly reproducible.  It should be recognized, though, that this model is not directly representative of atmospheric convection in which the buoyancy fluxes are thought not to be limited by the molecular diffusivities at the boundaries, and where re-evaporation can be important. We begin, in \secref{sec:model}, with the equations of motion and then, in \secref{sec:properties} and \secref{sec:nondimensionalization}, discuss the system's thermodynamic properties and the nondimensional parameters that govern it.  In \secref{sec:drizzle} we derive the drizzle solution, a state of no motion in which diffusion of moisture and buoyancy is exactly balanced by condensation and its release of latent heat, and which is the moist extension of the conductive state of \RB convection.   We explore the stability properties of that state and then, in \secref{sec:nonlinear}, we  describe some numerical solutions of the  full, nonlinear and generally time-dependent, system. Finally, in \secref{sec:discussion}, we conclude and discuss a pathway to more realism.

\section{The Model} \label{sec:model}

\subsection{Dry equations of motion} 
We first derive a set of dry Boussinesq ideal gas equations.  Our equations may be derived by adding an incompressibility ansatz to the anelastic equations found in \citet{Vallis17}, and a similar set of equations was presented by \citet{Spiegel_Veronis60} and discussed by \citet{Mahrt86}.  However, as we need to keep explicit track of both potential temperature and temperature for the moist extension, it is more transparent to proceed ab initio.  The main assumptions are that the perturbation density is smaller than the basic-state density and that the temperature profile of the basic state is nearly adiabatic. We consider a Cartesian domain with $z$ the vertical coordinate and write
\begin{equation}
	\label{dry.0} 
   	\rho = \ro + \delta \rho, \qquad p = p_0(z) + \delta p, 
\end{equation}
where $\rho$ is density, $p$ is pressure,  $\ro$ is a constant,  $\ddd {p_0} z \equiv - \ro g$, and $|\delta \rho| \ll \ro$, which is the fundamental Boussinesq ansatz. The momentum equations may then be written in the Boussinesq approximation as
\begin{subequations}
    \label{dry.1}
\begin{align}   
     \DD \ub   & = - \del \phi + \nu \del^2 \ub, \\
     \DD w  & = - \pp \phi z + b + \nu \del^2 w .
\end{align}
\end{subequations}  
Here, $\vb=(u,v,w)$ is the three-dimensional velocity, $\ub$ is the horizontal velocity $(u,v)$, $w$ is the vertical velocity, $\phi = \delta p/\ro$ and $b = -g \delta \rho/\ro$ is the buoyancy. The mass conservation equation takes the incompressible form
\begin{equation}
	\pp ux + \pp vy + \pp wz = 0.
\end{equation}

To obtain a thermodynamic equation we begin with  the first law of thermodynamics, written for an ideal gas and without approximation, as \citep[e.g.,][Chapter 1]{Vallis17}
\begin{equation}
	\label{dry.2} 
   	\cp \DD T - {1  \over \rho} \DD p = Q,
\end{equation}
where $Q$ is the heating, which includes diffusion. The second term on the left-hand side may be approximated by
\begin{equation}
	\label{dry.3} 
   {1 \over \rho} \DD p \approx \bfrac{w}{\ro} \dd {p_0} z = - w g, 
\end{equation}
and \eqref{dry.2} becomes
\begin{equation}
	\label{dry.3a} 
   	\DD{} \left(\cp T + gz  \right)  = Q,
\end{equation}
 or
\begin{equation}
	\label{dry.3aa} 
   	\DD{\theta}  = {Q \over \cp},
\end{equation}
where 
\begin{equation}
	\label{dry.3b} 
 \theta \equiv T + {g \over \cp} z
\end{equation}
is the potential temperature for this system, and $g/\cp$ is the `dry adiabatic'  lapse rate.

It remains to relate the buoyancy to the potential temperature. The density perturbations may be written as
\begin{equation}
	\label{dry.4} 
    \begin{split}
   	\delta \rho  & = \pp \rho T \delta T + \pp \rho z \delta z \\
                 & = \pp \rho T \delta T - \pp \rho T \pp Tz \delta z.       
    \end{split}             
\end{equation} 
We now assume (in common with most derivations of anelastic equations) that deviations from a neutral profile are small and that $\ppp T z \approx - g/\cp$. Equation \eqref{dry.4} becomes
\begin{equation}
	\label{dry.5} 
   	\delta \rho   = \pp \rho T \left( \delta T +   {g \over \cp} \delta z \right)  = - \frac \rho T \left( \delta T +   {g \over \cp} \delta z \right).            
\end{equation}
using the ideal gas equation. Using \eqref{dry.3b} the right-hand side of the above is just $-\rho/T$ multiplied by the perturbation potential temperature and, since $\rho = \ro$ and $T = T_0$ at lowest order, we have
\begin{equation}
	\label{dry.6} 
   	{\delta \rho \over \ro} \approx - {\delta \theta \over T_0} .
\end{equation}

The complete set of equations, including viscous and diffusive terms, are then
\begin{subequations}
    \label{dry.7}
\begin{align}   
     \DD \vb   & = - \del \phi + b \hat{\boldsymbol{z}} + \nu \del^2 \vb, \\
     \nabla \cdot \vb & = \pp ux + \pp vy + \pp wz = 0 , \\
     \DD b & =  \kappa \del^2 b + Q_b,   
\end{align}%
\label{dim_ns1}%
\end{subequations}%
where $Q_b$ represents the effects of non-diffusional heating on the buoyancy.
These equations are identical to those of a simple Boussinesq system with a linear equation of state. The temperature is related to the buoyancy by 
\begin{equation}
	\label{dry.8} 
   	T = T_0 + \delta T = \theta_0 + \delta \theta  - {g \over \cp} z, \qquad \text{or} \qquad
       \delta T = {\theta_0 b \over g} - {g \over \cp} z, 
\end{equation}
where $\theta_0 = T_0$ and $b = g \delta \theta/T_0$.

\subsection{Moist equations of motion}

Taking specific humidity $q$ (water vapour content divided by total mass of a parcel) as a materially conserved variable, except for diffusion and condensation, we have
\begin{equation}
	\label{moist.1} 
   	\DD q = \kappa_q \del^2 q + C,
\end{equation}
where evaporative effects are contained in the diffusion term and $C$ represents condensation. We assume that condensation occurs almost immediately upon saturation, reducing the value of $q$ to its saturated value. One may represent this process by
\begin{equation}
	\label{moist.2} 
   	C = -\frac{q - q_s}{\tau} \H(q - q_s)
\end{equation}
where $\H$ is the Heaviside function and $\tau$ is the timescale for condensation, which in the `fast condensation limit' is smaller than any other dynamical timescale in the system. All the condensate is then assumed to disappear from the system, and no liquid water is present.  We regard \eqref{moist.2} as a simple representation of the condensation process  and a means of avoiding complications with the uncertain microphysical processes employed in more complex models that retain the liquid and ice phases \citep{White_etal17, Zhao_etal16, Khain_etal15}.   An analogous fast autoconversion process between phases is sometimes used in these types of models.   Fast condensation may be regarded as a limiting form of a unary chemical reaction \citep{Pierrehumbert_etal07}; similar equations were also used by or appeared in the theoretical studies of \citet{Tsang_Vanneste17}, \citet{Sukhatme_Young11} and \citet{OGorman_Schneider06}, as well as in some idealized General Circulation Models \citep[e.g.,][]{Frierson_etal06}, and we refer the reader to these papers for more discussion.  It is by choice that we omit liquid and ice phases of water in order to make a closer connection to the Rayleigh--B\'enard problem, and this is one of the ways our model differs from cloud resolving models.

The value of $q_s$ is given by an approximation to the solution of the Clausius--Clapeyron equation for an ideal gas, as follows.  If the latent heat of condensation, $L$, is constant (in actuality $L$ for water varies by about 10\% between 0\dgc and 100\dgc) then the saturation vapour pressure, $e_s$, is given by
\begin{equation}
	\label{moist.3} 
   	e_s = e_0 \exp \left[ \frac{L}{R_v} \left(\frac{1}{T_0} - \frac{1}{T} \right) \right]. 
\end{equation}
where $R_v$ is the gas constant for water vapour and $e_0$ and $T_0$ are constants.  For small variations in absolute temperature \eqref{moist.3} approximates to
\begin{align}
	\label{moist.4} 
   	e_s = e_0 \exp\left( \frac{L (T - T_0) } {R_v T_0^2} \right) = e_0 \exp(\alpha \, \delta T),
\end{align}
where $\alpha = L /(R_v T_0^2) $ and $\delta T = T - T_0$. 

The specific humidity  is related to the vapour pressure, $e$, by
\begin{equation}
	\label{moist.4b} 
   	q = \frac {\epsilon e }{p - e} \approx \epsilon \frac{e_s}{p},
\end{equation}
 where $\epsilon$ is the ratio of the molar mass of water vapour (or other condensate) to that of dry air and is approximately 0.62 for water and air.  The approximation that $p \gg e$ is normally a very good one  for Earth's atmosphere.

The lowest order approximation to pressure in \eqref{moist.4b} is to take $p = p_0$ so that 
\begin{equation}
	\label{moist.5} 
   	q_s = q_0 \exp(\alpha \, \delta T) ,
\end{equation}
where $q_0 = \epsilon e_0/p_0$ is the value of the saturation specific humidity when $\delta T = 0$. More accurately we may write
\begin{equation}
	\label{moist.6} 
  p = p_0 \left( \frac{T}{\theta} \right)^{\tfrac{c_p}{R_d}}	\approx p_0 \left( \frac{T}{\theta_0} \right)^{\tfrac{c_p}{R_d}},
\end{equation}
assuming that $\theta\approx\theta_0$, and where $R_d$ is the gas constant of dry air. If we expand $T$ about a reference state, $T_0$, then a little manipulation leads to 
\begin{equation}
	\label{moist.7} 
  p = p_0 \exp \bfrac{\cp \delta T}{R_d T_0}	, 
\end{equation}
and thence, using \eqref{moist.4} and \eqref{moist.4b}, to 
\begin{equation} \label{ccrel}
q_s = q_0 \exp (\alpha' \delta T)
\end{equation}
where $\alpha' = \alpha - \cp/(R_d T_0) =  L /(R_v T_0^2) - \cp /(R_d T_0)$. This relation has the same form as \eqref{moist.5} and we henceforth drop the prime on $\alpha$. For water vapour, $L /(R_v T_0^2)$ is about three times larger than $\cp /(R_d T_0)$ at 273 K. We refer to \eqref{ccrel} and similar equations as Clausius--Clapeyron relations, and we henceforth refer to $\delta T $ as $T$.

The condensation of moisture provides a buoyancy source that can be determined from the first law of thermodynamics
\begin{equation}
	\label{moist.8} 
   L \, \Delta q = \cp \, \Delta T = {\cp  \theta_0 \over g} \Delta b ,
\end{equation}
or 
\begin{equation}
	\label{moist.9} 
    \gamma \Delta q = \Delta b ,
\end{equation}
where $\gamma = gL/(\cp  \theta_0)$ is a constant, so providing a non-zero $Q_b$ term on the right-hand side of the buoyancy equation (\ref{dry.7}c).

\subsection{Dimensional equations}

The full, dimensional, equations of motion are as follows. 
\begin{subequations}
    \label{dim.1}
\begin{align}   
     \DD \ub   & = - \del \phi + \nu \del^2 \ub, \\
     \DD w  & = - \pp \phi z + b + \nu \del^2 w , \\
     \DD b & = \gamma \dfrac{q - q_s}{\tau} \H(q-q_s)+ \kappa \del^2 b,\\
     \DD q & =  - \dfrac{q - q_s}{\tau} \H(q-q_s)+ \kappa_q \del^2 q, \\
     	\div \vb &\equiv \left(\pp ux + \pp vy + \pp wz \right) = 0. 
\end{align}
The temperature is obtained from the buoyancy by
\begin{equation}
 T = \dfrac{\theta_0}{g} b - \dfrac{gz}{c_p}.
\label{temp_vs_buoy}
\end{equation}
and the saturation humidity is given by
\begin{equation}
    q_s = q_0 \er^{\alpha T},
    \label{Clau-clap}
\end{equation} 
\end{subequations}   
where $q_0$ and $\alpha$ are constants.  The buoyancy here is not a function of the amount of water vapour in the system, which is appropriate if the system is sufficiently dilute. (Meteorologists account for this by use of a `virtual temperature', but here we ignore that effect.)   A useful measure of the degree of saturation of a parcel is the relative humidity, $r$, which here we define as $r = q/q_s(T)$. 

The only difference from the standard Boussinesq \RB equations is the presence of the moisture term, and because $q_s$ is a function of absolute temperature (not buoyancy), it is necessary to keep track of the local temperature via \eqref{temp_vs_buoy}.   The moisture term can be expected to be particularly important in updraughts: air that is rising will cool, become saturated and condense, releasing heat and increasing the buoyancy. On the other hand, descending air becomes warmer and less saturated and (absent diabatic effects) parcels typically follow the dry adiabatic lapse rate.  

The system (\ref{dim.1}\textit{a--g}) is a closed set of dimensional equations that need to be supplemented by boundary conditions on the variables. In this paper we take the variables to be periodic in the horizontal and generally impose no-slip, fixed temperature (and hence fixed buoyancy) and fixed specific humidity on the horizontal boundaries. If we set the specific humidity to be at its saturated value at the top and bottom boundary then typical boundary conditions are
\begin{subequations}
    \label{dim.5}
\begin{alignat}{2}
    b(z=0) &= 0,  \qquad & b(z=H) &= - \Delta b, \\
     T(z=0) &= 0,  \qquad & T(z=H) &= T_2 , \\
    q(z =0) &= q_0, \qquad & q(z=H) &= q_s(T_2).
\end{alignat}
\end{subequations}  
where $H$ is the top of the domain and  $T_2 = - \Delta b  (\theta_0/g)  - gH/\cp$. Alternatively, one might set $\ppp qz = 0$ at the top boundary, so that there is no source or sink of humidity at the top.  

Of the models mentioned in the introduction, the above system differs from the non-precipitating  models of \citet{Bretherton87} and \citet{Pauluis_Schumacher10} in its treatment of condensation and the equation of state  --- in our model water vapour is almost immediately removed from the system once saturation is reached, and consequently clouds in the usual sense do not form.  Our system has some similarities with the precipitating system of \citet{Hernandez-etal13}, but it is still simpler for we do not retain liquid water.  We also retain a temperature dependence in the Clausius--Clapeyron relation (in the Hernandez-Duenas system the saturated vapour pressure is just a function of height), and this is important both for the drizzle solution of \secref{sec:drizzle} and for the  nonlinear, time-dependent solutions.

\section{Thermodynamic and Conservation Properties} \label{sec:properties}
\subsection{Moist static energy and saturated lapse rate}
From (\ref{dim.1}c) and (\ref{dim.1}d) we obtain 
\begin{subequations}
\begin{equation}
	\label{cp.1} 
   	\DD{m} =  \kappa \del^2 b +  \gamma \kappa_q \del^2 q ,
\end{equation}
where
\begin{equation}
	\label{cp1b} 
   	m = b + \gamma q.
\end{equation}
\end{subequations}
In the absence of diffusion the quantity  $m$ is conserved on parcels, even in the presence of condensation, and is akin to a `moist static energy'  for this system. If we integrate over volume and time then, in a statistically steady state,  
\begin{equation}
	\label{cp.1b} 
   	 \int  \kappa \del b \cdot d \bm{\ell} = 
        - \int \gamma \kappa_q \del q \cdot d \bm{\ell} , 
\end{equation}
where the integral is over the boundary.  If there is condensation in the interior then there a buoyancy source in the interior, and there must be a buoyancy loss at the boundaries that balances the buoyancy gain due to condensation.  The loss of water due to condensation is balanced by moisture diffusion at the boundary, and \eqref{cp.1b} represents the balance between buoyancy loss and moisture gain at the boundaries. 

The lapse rate, $- \ppp T z$, corresponding to $\ppp m z = 0$ with $q = q_s$, is the so-called saturated adiabatic lapse rate, $\Gamma_T$ for this system, and this is easily calculated.  Using the definition of $b$ and $\gamma$, the state of $\ppp m z = 0$ corresponds to
\begin{equation}
	\label{cp.2} 
   	\pp T z + {g \over \cp} + {L \over \cp} \pp {q_s} T = 0.  
\end{equation}
Using the simplified Clausius--Clapeyron relation, \eqref{Clau-clap}, we obtain
\begin{equation}
 	\label{cp.3} 
    \Gamma_T = \frac{g/\cp} {1 + L \alpha q_s/\cp} 
             =  \frac{g/\cp} {1 + \alpha q_s \gamma \theta_0 /g} .
\end{equation}
We can also write this as a critical buoyancy gradient, $\Gamma_b = \ppp b z$, in which case we obtain 
\begin{equation}
 	\label{cp.4} 
    \Gamma_b =  \bfrac{g}{\theta_0} \left[\pp T z + {g\over \cp} \right]
             = \left[ { q_s \gamma \alpha g/\cp  \over 1 + q_s \gamma \alpha \theta_0 /g} \right].
\end{equation}
If $\gamma = 0$, or if the air is not saturated, then at criticality $\Gamma_b = 0$, as expected, but otherwise $\Gamma_b$ is a positive quantity. That is, the critical buoyancy lapse rate for convection is positive, and a gradient less than that of \eqref{cp.4} will be unstable (in the absence of diffusion) if the air is saturated. 

\subsection{Energy,  potential vorticity and boundary fluxes}
If the  latent heat of condensation is zero (i.e., if $\gamma = 0$) the momentum and buoyancy equations reduce to the conventional Boussinesq equations, and the moisture is just a passive scalar.  In the absence of viscosity and diffusion it is well known that the equations then conserve volume integrated energy,  $\int (\vb^2/2 -bz) \dV$. They also conserve potential vorticity on parcels -- meaning that $\DDD Q = 0$ where $ Q = \bm{\omega} \cdot \nabla b$, because the buoyancy term in the momentum equation is a function only of the advected variable $b$ itself. In the full moist ideal gas equations potential vorticity is not conserved because then the potential temperature is a function of moisture content as well as density and pressure and therefore cannot annihilate the solenoidal term \citep{Schubert_etal01, Vallis17}.

Neither energy nor potential vorticity are conserved in the fluid interior in the presence of condensation. The energy equation has a condensational source and is
\begin{equation}
	\label{cp.5} 
   	\DD{} \left(\tfrac12 \vb^2 - bz \right)  =  - z (Q_b  +  Q_d) - \epsilon .
\end{equation}
where $Q_b =(\gamma (q - q_s)/{\tau}) \H(q-q_s) $ is the heating due to moisture condensation (the first term on the right-hand side of (\ref{dim.1}c)),  $Q_d = \kappa \del^2 b$, and $\eps$ is the kinetic energy dissipation.  In a statistically steady state the left-hand side vanishes when integrated over the domain and
\begin{equation}
	\label{cp.6a} 
   \left< z (Q_b +  Q_d) \right>  =  - \langle \epsilon \rangle  . 
\end{equation}
Since the right-hand side is negative definite, the total heating,  ($Q_b + Q_d$), must be negatively correlated with the height, a known result \citep{Paparella_Young02}. Or, put informally, the heating must be below the cooling if kinetic energy is to be dissipated.  The consequence of this is that, with periodic horizontal boundary conditions and a kinetic-energy dissipating flow,  diffusion must carry buoyancy out of the fluid at the top boundary.

Finally, we note that in statistically steady state the total precipitation (namely the volume integral of the first term on the right-hand side of (\ref{dim.1}d) is (rather obviously) exactly balanced by moisture fluxes at the boundary. From (\ref{dim.1}c) the precipitation is also related to the boundary buoyancy fluxes. This type of result can provide constraints on how precipitation can evolve under climate change \citep[e.g.,][]{OGorman_etal12}, although we do not explore that issue here.

\section{Nondimensionalization} \label{sec:nondimensionalization}
\subsection{Diffusive scaling}
Here we nondimensionalize the system using the diffusivity and the height of the domain to define a timescale, as is common in \RB convection problems. In the appendix we present two other ways of proceeding, using buoyancy and moisture to provide time scales.  We scale all lengths $x,y,z$ with $H$, the depth of the domain, and we scale $q$ with $q_0$. 
The thermal diffusion time, $H^2/\kappa$ then scales time, and velocity scales as $\kappa/H$.  We  scale pressure with velocity squared, though in the Boussinesq formulation pressure is a Lagrange multiplier to ensure incompressibility and need not directly enter the dynamics.   We have a choice for scaling typical temperatures and buoyancy variables, either with the temperature or buoyancy drop across the layer, and because we shall investigate atmospheres which are stable (or marginally stable) to dry convection (so sometimes the buoyancy drop is zero) it is more useful to select the nondimensionalization with temperature drop. Thus, with hats denoting nondimensional variables and either an uppercase or a subscript $S$ or  $0$ denoting scaling values, we choose time ($t_s$), velocity ($U,W)$, buoyancy ($B$), moisture ($Q$) and pressure ($\Phi$) scales as, respectively, 
\begin{equation}
\begin{gathered}
\label{ndim.6}
     t_S = {H^2}\kappa, \qquad U = W =  \kappa / H ,  \qquad 
      B = {g \Delta T}/{\theta_0}, \\
      Q = q_0, \qquad  \Phi = U^2 = (\kappa/H)^2,  
\end{gathered}
\end{equation}
where $\Delta T$ is the temperature difference between bottom and top. The various variable may then be written as
\begin{equation}
\begin{gathered}
	\label{ndim.7}
   (x,y,z) =  (\xhat, \yhat, \zhat) H, \quad  (u,v,w) = (\uhat, \vhat, \hatw )  \kappa/H,  \\ 
    \quad b = \bhat \dfrac{g \Delta T}{\theta_0},  \quad T = \That \Delta T, \quad q = \qhat q_0, \quad  m = \mhat \dfrac{g \Delta T}{\theta_0}. 
\end{gathered}
\end{equation}
We also define $\tauhat = \tau \kappa / H^2$ to be the condensation time in units of the thermal diffusion time. 
With these choices the momentum equations  become 
\begin{align}
  \DDhat \ubhat & = - \del\phihat
     + \Pr \del^2 \ubhat.  \label{umom_scal1}\\
    \DDhat \hatw & = - \dfrac{\partial \phihat}{\partial \zhat}
          + \Pr \Ra \, \bhat + \Pr \del^2 \hatw.  
          \label{wmom_scal1}
\end{align}
The buoyancy and moisture equations may be written as
\begin{align}
\DDhat \bhat & = \gammahat \dfrac{\qhat - \qhat_s}{\tauhat} \H(\qhat - \qhat_s)+ \del^2 \bhat  \label{buoy_scal1}
\\ 
\DD \qhat & =  \dfrac{\qhat _s- \qhat}{\tauhat} \H(\qhat - \qhat_s) + \Sm \del^2 \qhat.  \label{q_scal1}
\end{align}
The nondimensional numbers  are:
\begin{equation}
	\label{ndim.12}
	\Ra = \dfrac{\Delta T\, g H^3}{\theta_0 \kappa \nu}, \qquad \Pr = \dfrac{\nu}{\kappa},
	\qquad 
	\Sm =   {\kappa_q \over \kappa}, \qquad \gammahat = \gamma\bfrac{q_0 \theta_0}{g \Delta T} = {L q_0 \over c_p \Delta T} . 
\end{equation}
Here, $\Ra$ is the familiar  Rayleigh number, $\Pr$ the Prandtl number, $\Sm$ is a ratio of diffusivities, 
and $\gammahat$ is the nondimensional value of $\gamma$, which in turn is a measure of the importance of the latent heat of condensation.  The nondimensional moist static energy is given by $\mhat = \bhat + \gammahat \qhat$. 

The `physics' equations are 
\begin{equation}
    \label{ndim.13}
\eqnab
    q_s = q_0\exp(\alpha T),  \qquad T = \dfrac{\theta_0}{g} b - \dfrac{g}{c_p} z\
\end{equation}  
With the above nondimensionalization these equations become
\begin{equation} \eqnab
 \qhat_s = \exp(\alphahat \That),
 \qquad
 \That = \bhat - \betahat \zhat.
 \label{Tq_scal1}
\end{equation}
where 
\begin{equation}
	\label{ndim.14} \eqnab
\alphahat = \alpha \Delta T = {L \Delta T \over R_v T_0^2}, \qquad
 \betahat = { g H \over c_p \Delta T} . 
\end{equation}

 Two other ways to nondimensionalize the system are given in the appendix, one using the buoyancy difference across the layer to scale time, and the other using the buoyancy created by condensation to scale time. The nondimensional numbers that arise are combinations of those above,  and we just note one of them here, the condensational Rayleigh number:
\begin{equation}
\label{Rgamma1}
\Rg = \dfrac{ g L H^3 q_0}{c_p \theta_0 \kappa \nu}  = \Ra \bfrac{q_0 L}{\Delta T \cp}  = \Ra \gammahat.
\end{equation}

\subsection{Nondimensional lapse rates}
\newcommand{\Gammahat}{\widehat \Gamma}
The nondimensional dry adiabatic lapse rate is given by
\begin{equation}
	\label{lr.1} 
   	\Gammahat_d =  \frac{g }{\cp} {\Delta T \over g}  = \betahat ,
\end{equation}
consistent with (\ref{Tq_scal1}b).  The nondimensional saturated adiabatic lapse rate for temperature is given by 
\begin{equation}
 	\label{cpnondim.3} 
    \Gammahat_T =  {\Delta T \over g} \frac{g/\cp} {1 + L \alpha q_s/\cp}  = \frac{\betahat}{1 + \gammahat \alphahat \qhat_s} ,
\end{equation}
and the corresponding  lapse rate for buoyancy is given by
\begin{equation}
 	\label{cp.4b} 
    \Gammahat_b =  \frac{M \betahat \alphahat \qhat_s }{ 1 + \gammahat \alphahat \qhat_s} .
\end{equation}
The deviation from the dry values is proportional to the product $\gammahat\alphahat$, which is roughly proportional to the square of the latent heat of vapourization.

\subsection{Parameter values}
 Aside from the ratio of the moisture diffusivity to the viscosity or thermal diffusivity, the new nondimensional numbers in the system (compared to the \RB model) are:
\begin{enumerate}
  \item The parameter $\gammahat$ is the ratio of the buoyancy effect due to the heat release by condensation to the dry effect, 
  namely $\gammahat = L q_0/(\cp \Delta T)$. The condensation Rayleigh number,
  $\Rg = { g L H^3 q_0}/(c_p \theta_0 \kappa \nu)$, captures the same effect and $\gammahat = \Rg/\Ra$.
  \item The dry adiabatic lapse rate, $\betahat = g H/(\cp \Delta T)$,  derived from the dimensional value $g/\cp$.
  \item The parameter $\alphahat$ determining the exponential growth of saturation specific humidity, $\alphahat = L \Delta T/(R_v T_0^2)$.  The corresponding dimensional parameter is $L/(R_v T_o^2)$.   
\end{enumerate}

\newcommand{\s}{\,\text{s}}  \newcommand{\km}{\,\text{km}} 
\newcommand{\J}{\,\text{J}} \newcommand{\kg}{\,\text{kg}} \newcommand{\Kv}{\,\text{K}\xspace} \newcommand{\m}{\,\text{m}}
In Earth's atmosphere approximate dimensional values are 
$L = 2.5\eten 6 $\J/kg, $\cp = 1004 \J\kg^{-1} \Kv^{-1}$,  $\cp = 718 \J\kg^{-1} \Kv^{-1}$,
$R = 287\J\kg^{-1} \Kv^{-1}$, $R_v = 462  \J\kg^{-1} \Kv^{-1}$, $\epsilon = R_d/R = 0.622$, $e_0 = 611$\,Pa,  $p_0 = 10^5$ Pa, $q_0 = \epsilon e_0/p_0 = 3.8 \eten{-3}$,  $H = 10^4\m$,  $\Delta T= 50 \Kv$, $g = 9.8\m\s^{-2}$,  $\theta_0 = 300\Kv$,  $B = g \Delta T/\theta_0 \approx 1.6 \m^2\s^{-2}$, 
 $\nu = 1.5\eten{-5}\m^2\s^{-1} $,   $\kappa = 2.1\eten{-5}\m^2\s^{-1}$, $\kappa_q = 2.8\eten{-5}\m^2\s^{-1}$. 
Typical values of the nondimensional parameters are then
\begin{gather}
\label{nondims2}
  Pr  = \frac \nu \kappa \approx  0.7,  \quad 
 \Ra  = \frac{g \Delta T H^3 } {\theta_0 \nu \kappa }  \approx  5.4 \times10^{21},  \quad
   \Rg =    { g L H^3 q_0 \over (c_p \theta_0 \kappa \nu)  } \approx 1.0 \eten{21} , \\
     \qquad   \gammahat = { L q_0 \over \cp \Delta T}  =  {\Rg \over \Ra } \approx 0.19 ,  \quad
    \alphahat ={ L \Delta T \over R_v T_0^2} \approx 3.0, 
    \quad \betahat = {g H \over (\cp \Delta T)  }\approx 1.95
\end{gather}

Apart from the Rayleigh numbers these quantities are all of order unity in Earth's atmosphere.  The fact that $\gammahat \sim 1$ means that water condensation is, roughly, as important a source of buoyancy as is the temperature gradient between the ground and the tropopause (which ultimately comes from radiation). As $\alphahat \sim 1$ the water vapour content falls off significantly over the depth of the domain.  Because $\gammahat \alphahat \sim 1$ the moist adiabatic lapse rate is typically noticeably less than the dry adiabatic lapse rate (and detailed calculations indicate it varies by a factor of 0.3 to 1.0 of the dry adiabatic lapse rate). The fact that $\betahat \sim1$  indicates that temperature significantly differs from potential temperature over the depth of the troposphere, but is of the same order of magnitude. 

For comparison, consider also the values for Jupiter and Titan, two other atmospheres with condensates in the Solar System.  On Jupiter water vapour is a condensate but the predominant gas is hydrogen (so that $\epsilon$ and the heat capacities differ),  and gravity is  larger.   In Jupiter's weather layer (at a pressure of 3 bars, say) we might take  $\cp = 1.4\eten4 \J\kg^{-1} \Kv^{-1}$, $H = 100\km$,  $p_0 = 3\eten 5$\,Pa,  $g = 50\m\s^{-2}$, $\epsilon =  9$, $q_0 = \epsilon e_0/p_0 = 1.8\eten{-2}$,  $\Delta T = 100\Kv$,  giving  $\gammahat = 0.03$ and $\alphahat = 6$ . The value of $\gammahat$ is  smaller than that for Earth because of the higher value of pressure chosen (3 bars instead of 1), and the larger temperature difference suggesting that water condensation is not a primary factor over the whole Jovian troposphere (meaning the weather layer below the stratosphere),  but may still be important in the lower troposphere.   The value of $\gammahat \alphahat$ is about  0.2, indicating a moist adiabatic lapse rate somewhat less than the dry one, but not significantly so.  The value of $\betahat$ is about $3.6$ (for $H =100\km$) indicating that the temperature differs from the potential temperature over the depth of the troposphere. 

On Titan, which has a surface temperature of about 100\Kv, nitrogen is the main component of the dry atmosphere and  methane is the main condensate, with concentrations varying between about 1\% and 5\%.  The latent heat of condensation for methane is $ L \approx  4.8\eten 5 \J\kg^{-1}$ and its gas constant is $R_v = 518\J\kg^{-1}\Kv^{-1}$. The heat capacity of its nitrogen atmosphere is about the same as air on Earth so that $\cp \approx 10^3\J\kg{-1}\Kv^{-1}$, and $g\approx 1.35\m\s^{-2}$ and there is an estimated temperature drop of about 25\Kv over a tropospheric depth of 40\km, with a surface pressure of roughly $10^5$\,Pa, and a representative value of $q_0$  for methane is $10^{-2}$.   Typical values are then $\gammahat = 0.2$, $\alphahat  = 2.3$, not dissimilar to those for  Earth, suggesting that condensation is an important source of buoyancy and that the moist adiabatic lapse rate  will be noticeably smaller than the dry one, as detailed calculations confirm \citep{Mitchell_Lora16}.  We also find $\betahat = 1.6$, similar to Earth's value.

\subsection{A note on parameter independence and thermodynamic consistency}

The nondimensional parameters $\gammahat, \alphahat$ and $\betahat$ contain different combinations of various physical constants. This means that, even though $L$ appears in both $\gammahat$ and $\alphahat$, and $\cp$ appears in both $\gammahat$ and $\betahat$, the values of the nondimensional parameters may be varied independently without thermodynamic inconsistency.  The  value of $L$ is related to the difference in the heat capacities of the condensate in its liquid and vapour forms,  $\cp^l$ and $\cp^v$ respectively, but this is not the only  factor  determining $L$. (To a good approximation, $L = L_0 + (\cp^v - \cp^l)T$,  but the $L_0$ term usually dominates and so $L$ is almost constant and independent of the heat capacity of the condensate.)   Thus, although the gas constant of  the vapour, $R_v$, does appear in the expression for $\alphahat$ along with $L$,  $\alphahat$ is effectively an independent parameter, meaning that it may be varied independently of $\gammahat$ (and of $\betahat$).  Varying the parameters does imply a variation in the physical nature of the condensate and the dry air, and, for example, if one wished to explore the effect of a reduction of the latent heat alone one would need to change $\gammahat$ and $\alphahat$ consistently.   If one wished to  transition continuously to the dry system then a simple way would be to take $\gammahat \to 0$ and $ \alphahat \to 0$, noting that  $\alphahat$ has no effect when $\gammahat=0$. 

In the rest of the paper we drop the hats on the nondimensional variables and the dimensional/nondimensional distinction will be apparent by context. Also, unless explicitly noted, we take $\Pr = S_m = 1$ and we  take $\tau$ (formerly $\tauhat$) to be $5 \times 10^{-5}$.

\section{The Drizzle Solution} \label{sec:drizzle}

The classical \RB problem has a diffusive solution of no motion and a linear gradient of buoyancy connecting the upper and lower boundaries. The \rb problem has an analogous diffusive solution, but moisture (as well as buoyancy) now diffuses from the lower boundary, condenses and falls out as drizzle, at a rate determined by the rate of diffusion.\footnote{Meteorologists define drizzle as light precipitation with small water droplets, usually less than 0.5 mm in diameter \citep[]{AMSGlossary}. Here we have no droplet size, but the precipitation is very light compared to our convecting solutions.} Furthermore, the condensation occurs at saturation, and the temperature, and hence the value of the saturation humidity, is determined by the release of latent heat.  The upshot of this is that we have a nonlinear problem even in the absence of motion, and we now find its solution.

 \subsection{Saturated boundaries}
 
Let us calculate a time-independent basic state solution to the non-dimensional equations~\eqref{umom_scal1}, \eqref{wmom_scal1}, \eqref{buoy_scal1}, \eqref{q_scal1},  together with the physics equations, \eqref{Tq_scal1}.   In the case of no motion,  horizontal homogeneity and a saturated domain the thermodynamic and moisture equations are, with $S_m = 1$ and a diffusive nondimensionalization, 
\begin{subequations}
	\label{drz.0} 
\begin{align}
   \dd[2]b z & =  - \gamma \frac{q - q_s}{\tau}, \\
   \dd[2] q z & =  \frac{q - q_s}{\tau}, 
\end{align}
\end{subequations}
Evidently there is a solution of the moist static energy,  $m = b + \gamma q$ that obeys
\begin{equation}
\dd[2]m z = 0,
\end{equation}
yielding a linear dependence of $m$ with height, i.e.,  $m = P + Qz$, regardless of the presence or otherwise of condensation, where the constants $P$ and $Q$ are determined by the boundary conditions.  If we specify that $b=b_1$ and $b = b_2$ at the lower and upper boundaries, respectively, or equivalently $T_1 = b_1$ and $T_2 = b_2 - \beta$,  and that these boundaries are saturated with values $q_1$ and $q_2$ respectively, then,  at $z= 0$, 
\begin{equation}
\label{peqn}
     P = b_1 + \gamma q_1 = T_1 + \gamma q_1 = T_1 + \gamma \exp(\alpha T_1).
\end{equation}
At the top, $z=1$,  we have $b_2 + \gamma q_2 = P + Q $, whence
\begin{equation}
\label{qeqn}
    Q = [(b_2 - b_1) + \gamma (q_2 - q_1) ] = \beta + T_2 - T_1 + \gamma (\exp(\alpha T_2) - \exp (\alpha T_1) )  .
\end{equation}
The solution then satisfies $ b + \gamma q =  P + Q z $
with $P$ and $Q$ as above.  Now, for small $\tau$ and provided that moisture is converging everywhere,  the humidity must everywhere be very close to its saturated value. That is,  $q  =  q_s + \tau q' + \mathcal O(\tau^2)$
and  therefore 
 \begin{equation}
 b + \gamma q_s  =  P + Q z + {\cal O}(\tau) .
 \end{equation}
Using the Clausius--Clapeyron relation, \eqref{Tq_scal1}, for $q_s$ then gives, for small $\tau$,
\begin{equation}
	\label{drz.n} 
P +  Q z = b + \gamma \exp(\alpha (b -\beta z)) , 
\end{equation}
or equivalently
\begin{equation}
	\label{drz.n2} 
P +  Q z = T + \beta z + \gamma \exp(\alpha T) . 
\end{equation}
This algebraic equation for $b$ or $T$  may be solved via an iterative root-finding algorithm, or analytically by use of the Lambert-W function, which is the function that satisfies the equation $\mathrm{W}(\chi) \exp(\mathrm{W}(\chi)) = \chi$ for any $\chi$.  Explicitly, the exact solution of \eqref{drz.n2} is given by 
\begin{equation}
	\label{drz.lamb1} 
   	T(z)  = C  - \frac{ \mathrm{W}(\alpha \gamma \exp(\alpha C)) }{\alpha} ~ , 
\end{equation}
where $C = P + (Q - \beta) z$ with $P$ and $Q$ given by \eqref{peqn} and \eqref{qeqn}.  As soon as $T$ is known then $q_s$ is given using (\ref{ndim.13}a).  From \eqref{drz.n} we also see that
\begin{align}
	\label{drz.check} 
   	\dd[2] b z &= - \gamma \alpha^2 \beta^2 \exp(\alpha (b - \beta z))  <0 , \\ 
      \dd[2] q z & =  \alpha^2 \beta^2 \exp(\alpha (b - \beta z))  > 0 .
\end{align}
That is to say, the solution for $q$ is convex and moisture does indeed converge to every point in the domain, meaning that the solution must be at least saturated everywhere, consistent with the assumption.

\begin{figure}
 \centering
   \includegraphics[width=0.8\textwidth]{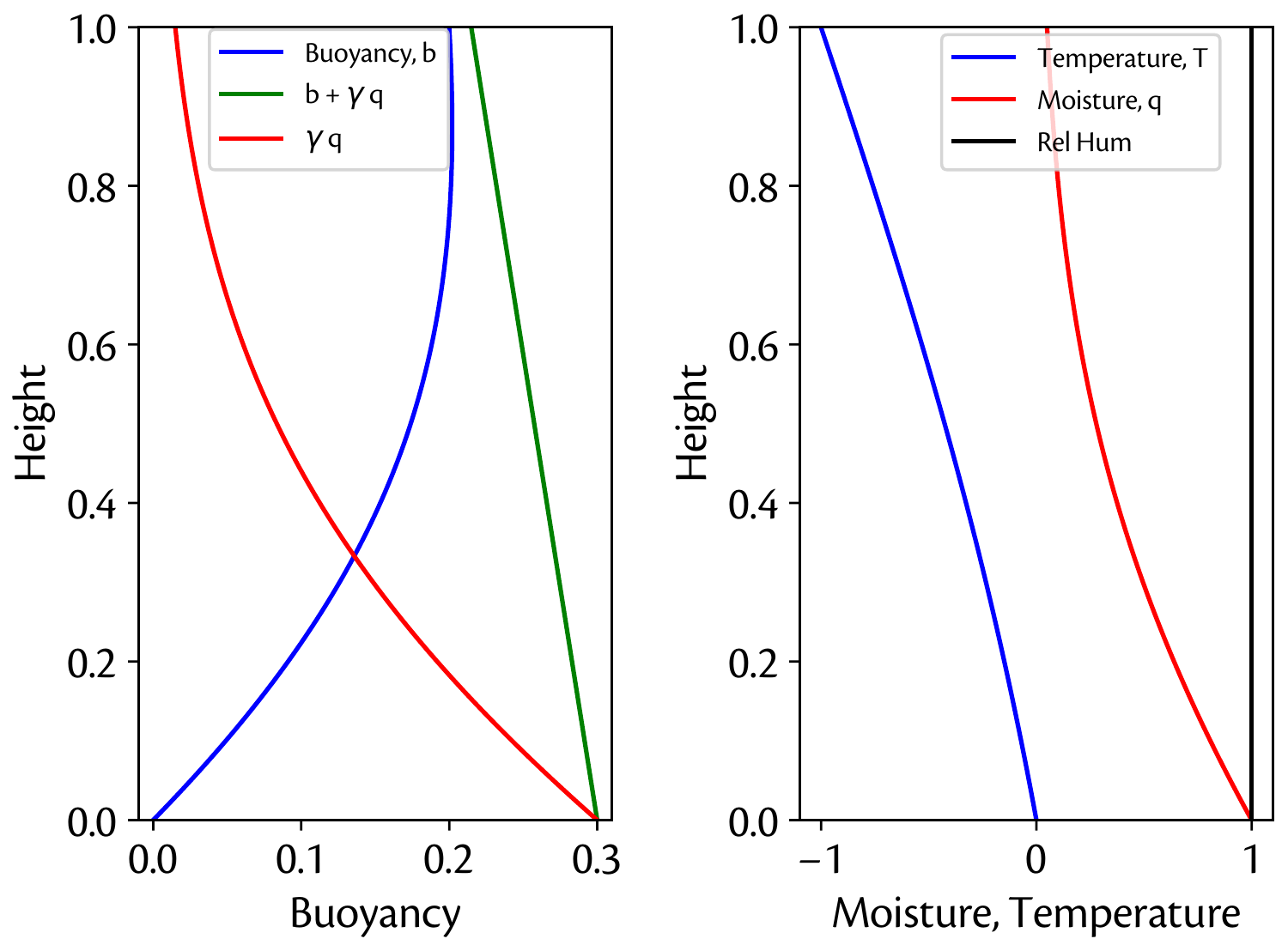} 
  \caption{A nondimensional drizzle solution with saturated upper and lower boundaries. The boundary conditions are that $b = 0$ at $z = 0$ and $b = 0.2$ at $z=1$ and $q = q_s$ at both boundaries, thus giving the values of $m$ ( $= b + \gamma q)$ at the boundaries shown. These solutions were obtained with $\gamma = 0.3$, $\alpha = 3$ and $\beta = 1.2$. }
\label{fig:drizzle-sat}
  \end{figure} 
  
The temperature and humidity profiles do not  depend on the value of the diffusivity. However, the rate of condensation (i.e., the  rainfall) does.  To see this explicitly we use the diffusion-based nondimensionalization and temporarily restore the hats on  nondimensional variables.  Setting the left-hand sides of \eqref{buoy_scal1} and \eqref{q_scal1} to zero  and using $ \qhat = \qhat_s + \tauhat \widehat{q'}$ gives 
\begin{equation}
	\label{drz.rain1}  \eqnab
   	  \gammahat \widehat {q'}  = -    \dd[2] \bhat \zhat, \qquad \text{and} \qquad
 \widehat {q'}  =   \dd [2] {\qhat_s }\zhat ,
\end{equation}
The dimensional value of the perturbation buoyancy, $q'$, is obtained from
\begin{equation}
	\label{drz.rain3} 
   	q'  = \tauhat \widehat{q'} q_0 = \frac{\kappa}{H^2}   \tau \widehat{q'} q_0 ,
\end{equation}
so that, re-dimensionalizing (\ref{drz.rain1}b), we obtain
\begin{equation}
	\label{drz.rain4} 
\frac{H^2}{\kappa q_0}   	{q'  \over \tau}	=     \dd [2] {q_s} z \frac{H^2}{q_0} , 
       \qquad \text{or} \qquad 	{q'  \over \tau} = \kappa \dd[2]{q_s} z . 
\end{equation}
This expression gives the condensation rate, since $\kappa \ddd[2]{q_s} z$ is the moisture convergence. Thus, the fields of buoyancy and moisture themselves do not depend on the diffusivity but the rate of condensation does.  Although this result may seem obvious from the form of the dimensional equations,  it is only when $\tauhat \ll 1$ (nondimensionally) or $\tau \ll H^2/\kappa$ (dimensionally) that it is valid, because only then is $q'$ small and $q \approx q_s$. 

Figure~\ref{fig:drizzle-sat} shows various  profiles for buoyancy, temperature, humidity and $m(z)$ for a typical set of non-dimensional parameters. For this set of parameters, which is stable to dry convection, the buoyancy increases with height in the lower part of the domain, but decreases slightly in the top part of the domain. The variable $m(z)$ is linearly decreasing with height, implying that the solution will be unstable to moist convection in the diffusion-free limit.  Finally, we remark that, at low Rayleigh number, numerical solutions of the nonlinear, time-dependent equations do converge to the drizzle solution. 

\subsubsection{The linear drizzle problem}
The drizzle problem may be further simplified by linearizing the solution to  the Clausius--Clapeyron equation, so that $ q_s = (1 + \alpha T)  = (1 + \alpha(b - \beta z))$, but this leads to an uninteresting case.  Suppose that the boundaries are saturated, with a higher temperature at the bottom. The solution to \eqref{drz.0} is then just a linear profile of both $b$ and $q$ between lower and upper boundaries.  Although the domain is saturated everywhere there is no convergence of moisture, and no condensation, because $\ddd[2]{q_s} z = 0$ everywhere. The moisture content of the fluid must increase faster than linearly with temperature  in order to produce condensation, as with the exponential relation \eqref{Clau-clap} that arises from the Clausius--Clapeyron equation.

\subsection{Drizzle solution with unsaturated boundaries}

In geophysical settings it is common to suppose that the lower boundary condition on moisture does not correspond to saturation; that is, the boundary -- for example moist soil -- is  damp but not saturated;  there will then be a layer of unsaturated air adjacent to the boundary. In the unsaturated layer the buoyancy and the moisture diffuse independently and each has a linear profile, but if the temperature falls off sufficiently rapidly with height then the water vapour will become saturated at some level, above which a solution similar to that of \eqref{drz.n} holds.  

\begin{figure}
\centering
\includegraphics[width=0.8\textwidth]{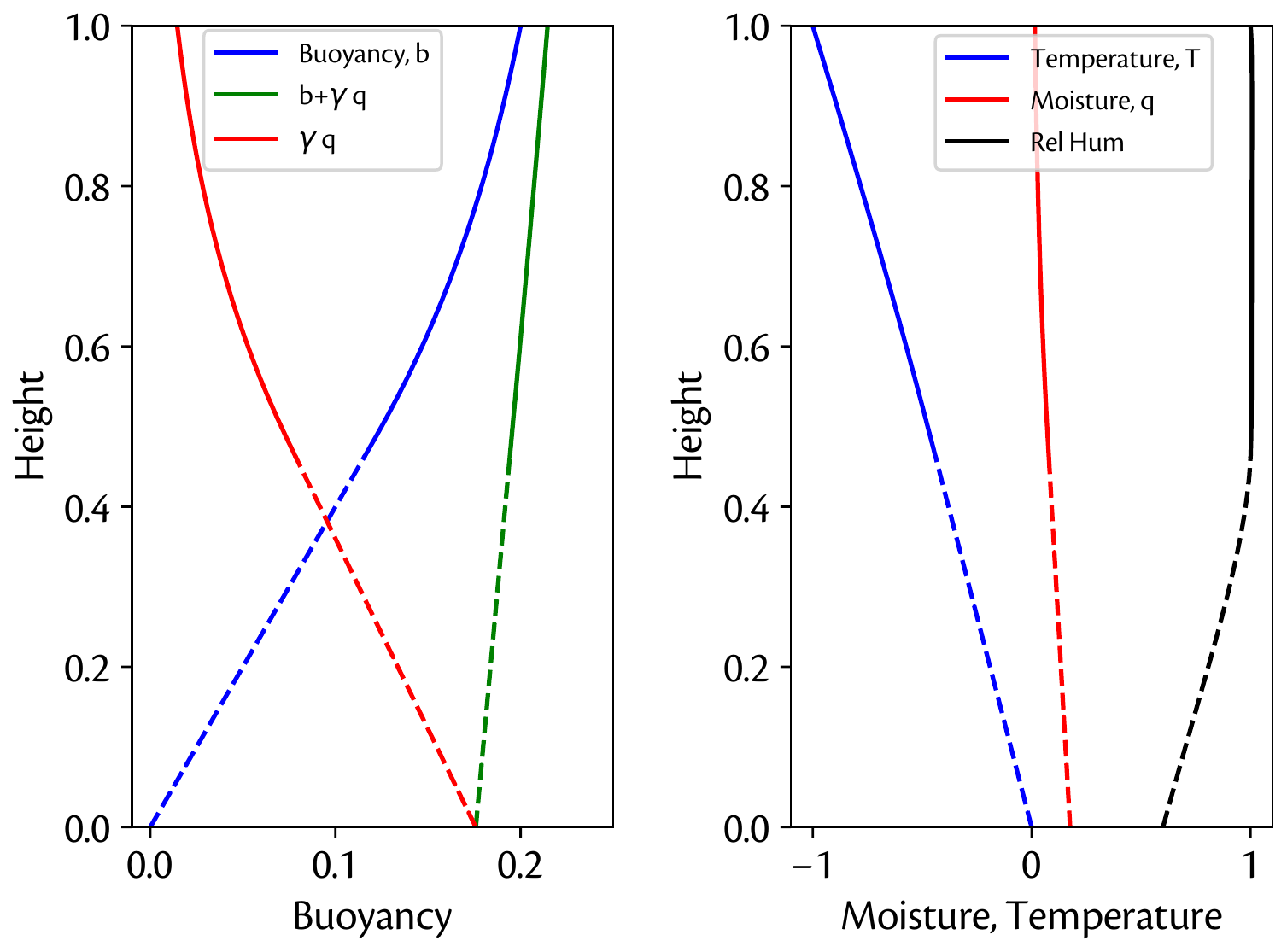} 
  \caption{Drizzle solution with an unsaturated lower boundary, with relative humidity ($q/q_s$) of 0.5. Other parameters are the same as in Fig. \ref{fig:drizzle-sat}. The dashed lines show the linear profiles of moisture and buoyancy up to a height of about 0.45, above which the solution is saturated.}
\label{fig:drizzle-unsat}
  \end{figure} 
  
Suppose that, as before, the boundary conditions at the bottom and top of the domain are ($b_1, q_1$)  and ($b_2, q_2$) respectively, but that $q_1$ is less than the saturated value.  The solution will be unsaturated below some height $z_c$ at which condensation begins, and for $0   < z < z_c$ the profiles of both $b$ and $q$ are linear in $z$, with no condensation.
At $z_c$ the humidity takes its saturated value at $z_c$ and therefore $q_c = q_s(b_c) = \exp[\alpha(b_c - \beta z_c)]$. 
Above $q_c$ region is completely saturated and the calculation proceeds, in principle, as in the previous subsection, but with different boundary conditions and over the smaller domain.  The height $z_c$ is determined by requiring that 
the first derivatives of temperature and moisture (as well as the values themselves) are continuous at the level of saturation, and this is most easily done numerically. That the first derivative should be continuous can be seen by integrating the governing equation across the condensation level giving
\begin{equation}
	\label{gov;drizzle1} 
   \int_{z_c-\eps}^{z_c+\eps} \!\! 	\kappa \pp[2] q z  \, \dz =  \int_{z_c-\eps}^{z_c+\eps}  \H(q - q_s) \frac{(q-q_s)}{\tau}  \,\dz .
\end{equation}
 The right-hand side goes to zero as $\eps \to 0$, even with the Heaviside function, and hence $\ppp q z$ remains continuous, and similarly for temperature and buoyancy. 

 If the lower boundary is chosen to be saturated the solution is that described previously, whereas if the lower boundary is sufficiently dry then there is no condensation and buoyancy and moisture both vary linearly from bottom to top. 
 A typical intermediate solution is illustrated in \figref{fig:drizzle-unsat} with relative humidity equal to 0.6 at the lower boundary.  Here the unsaturated region extends from the bottom to about $z = 0.45$ over which the relative humidity increases monotonically until the air is saturated, above which the solution has a shape similar to that in \figref{fig:drizzle-sat}, with the fields and their first derivatives remaining continuous throughout.

\subsection{Stability of the drizzle solution}

\begin{figure}
 \centering
 \includegraphics[width=\textwidth]{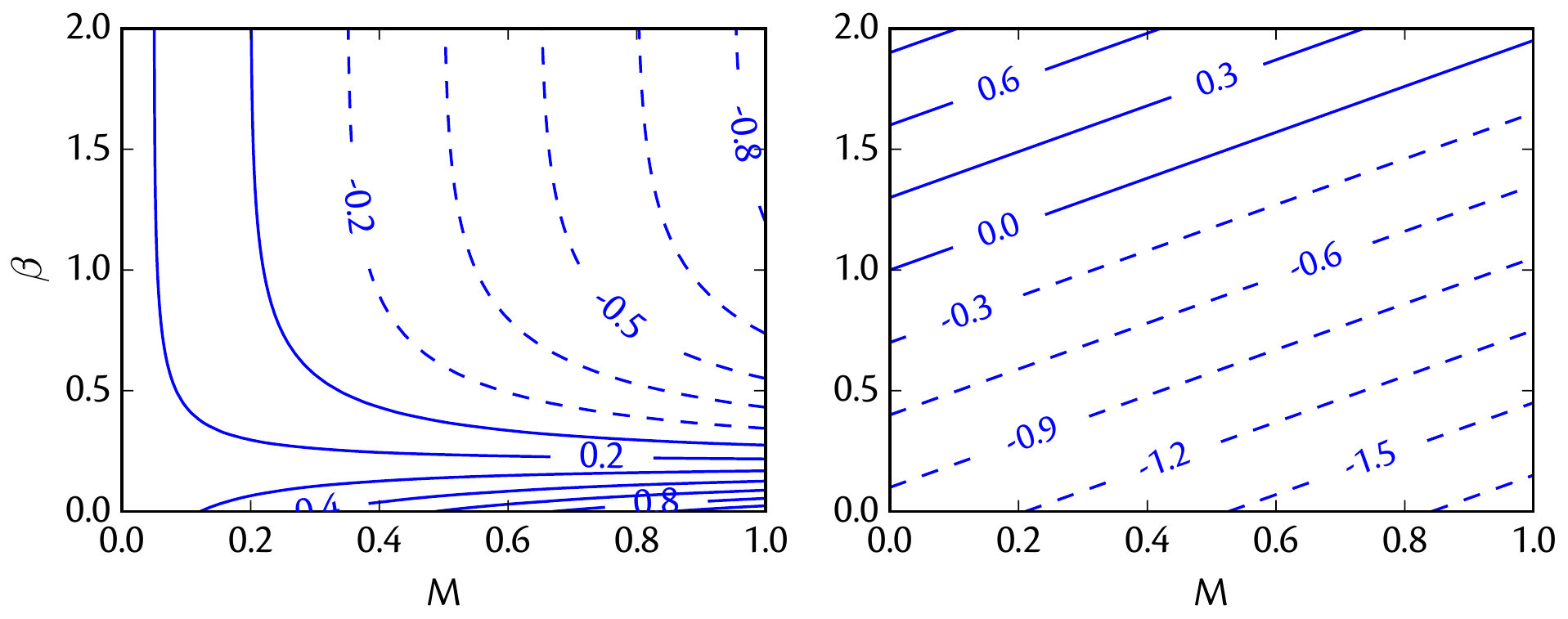}\\
  \caption{Contour plot of the value of the vertical gradient of $m(z) = b+ \gamma q_s$ (i.e., $\ppp mz$) in the drizzle solution as a function of $\beta$ and $\gamma$, with $\alpha = 3$. The left plot imposes values of $b(0) = 0$ and $b(1) = 0.2$ with $T = b - \beta z$. The right plot has $T(0) = 0$ and $T(1) = -1$, with $b = T + \beta z$. The boundary values of $q_s$ are calculated given the appropriate value of $T$.  Convective instability is possible in the region of dashed lines.}
\label{fig:drizz_cont}
 \end{figure} 

In the absence of diffusion the stability of the drizzle solutions is related to the sign of the gradient of the linear function $m(z)$. Figure~\ref{fig:drizz_cont} gives the sign of this gradient as a function of the non-dimensional parameters $\gamma$ and $\beta$ for the problem with saturated boundaries.  As expected when $\gamma=0$ (and moisture plays no role) stability is guaranteed if $\beta>1$ in the problem with specified temperature at the boundaries (right panel), and as $\gamma$ is increased the critical $\beta$ for non-diffusive stability increases linearly with $\gamma$. 

\begin{figure}
\centering
\includegraphics[width=0.65\textwidth]{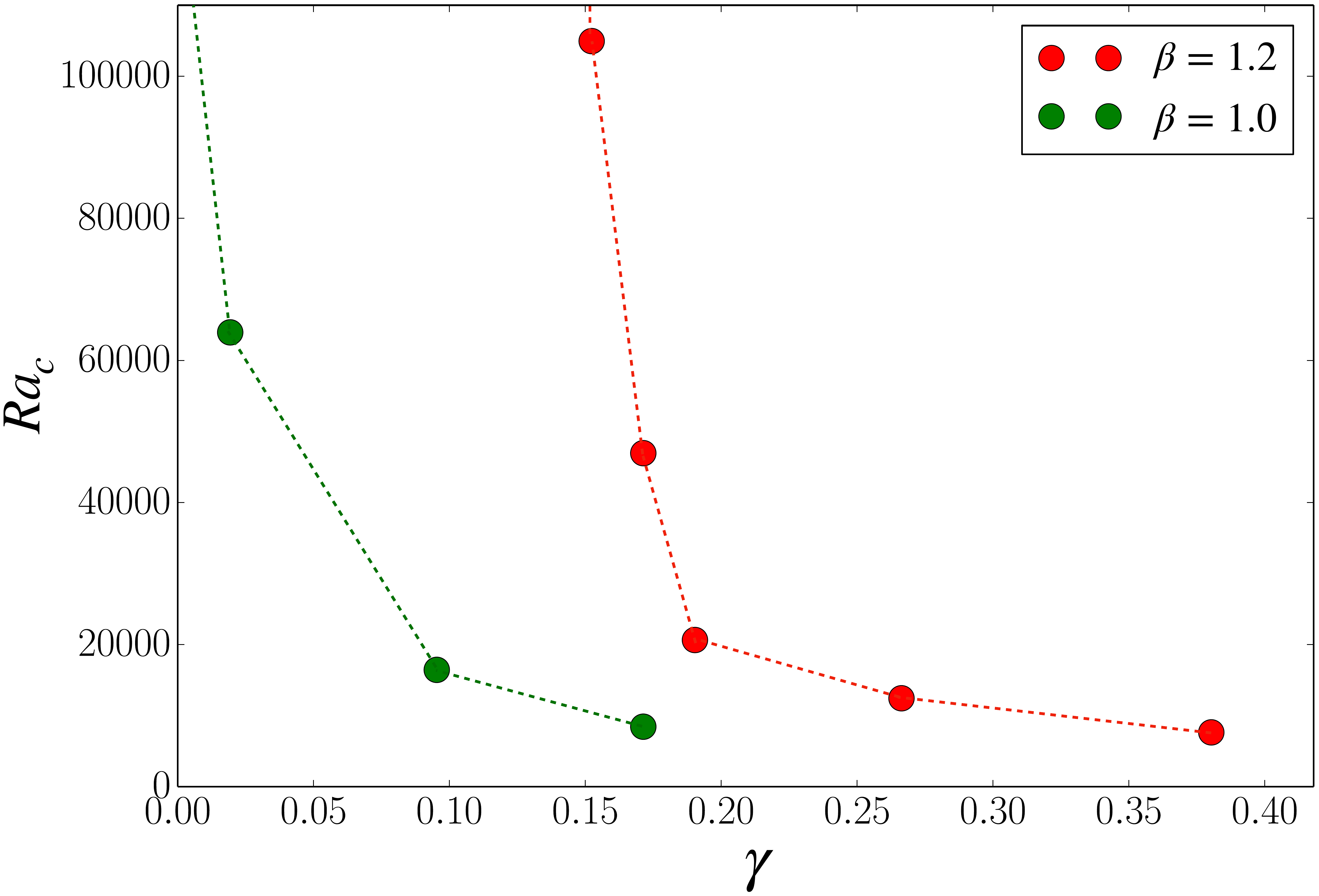} 
  \caption{Stability of the linear drizzle solution. Graph of $Ra_c$ versus $\gamma$ for the drizzle solution at $\beta=1$ (green) and $\beta=1.2$ (red).  The  dots mark the numerically determined stability boundary, with instability occurring above and to the right of the dots.}
  \label{fig:drizzle-stab}
\end{figure}

In the presence of diffusion, the stability of the drizzle solution is determined by $\gamma$, $\beta$, $Ra$ and $\alpha$. Even for cases where instability is allowed in the non-diffusive case, instability in the diffusive cases only occurs for sufficiently high $Ra$. The critical $Ra_c$ may be calculated by timestepping from a slightly perturbed numerically calculated drizzle solution in a large domain followed by a bisection method. (The method thus strictly finds the stability with respect to small but not infinitesimal perturbations.)  Figure~\ref{fig:drizzle-stab} shows the critical Rayleigh number $Ra_c$ as a function of $\gamma$ for two different choices of $\beta$:   $\beta=1.0$ corresponds to a neutrally dry marginally stable atmosphere and $\beta=1.2$ corresponds to a stable dry atmosphere, for the case with saturated boundaries.  We see that $Ra_c$ approaches an asymptote at the non-diffusive threshold ($\Ra \to \infty$), and as $\gamma$ is increased the critical Rayleigh number decreases.  For $\gamma = 0$ the problem is that of dry convection and the case with $\beta = 1.2$ is then always stable. The case with $\beta = 1.0$ and $\gamma=0$ will be stabilized by any non-zero diffusivity, and is neutrally stable in the infinite Rayleigh number limit. 

\section{Nonlinear Results: Initial Value Problems and Statistical Equilibration} \label{sec:nonlinear} 

In this section we discuss various results obtained with the full, time-dependent and nonlinear, system.   
Specifically, we solve nondimensional nonlinear equations ~(\ref{umom_scal1}), (\ref{wmom_scal1}), (\ref{buoy_scal1}), (\ref{q_scal1}), together with the physics equations~(\eqref{Tq_scal1}),  in two dimensions, with an aspect ratio of 20 and $\Pr = S_m = 1$.  In all experiments we set the atmosphere to be stable to dry convection, with $\beta = 1.2$, and set $\alpha=3.0$.  We first describe a sequence of experiments in which the Rayleigh number is varied, with all other parameters fixed, and then describe experiments with the parameter $\gamma$ varying. The equations are solved numerically using the flexible spectral solver Dedalus. We use Fourier series in the horizontal, Chebyshev polynomials in the vertical, and third-order semi-implicit backward time-differentiation. Heaviside functions are evaluated by $\H(\xi) \approx (1+\tanh k \xi)/2$, with $k=10^5$.  Flows with positive and negative vertical velocity are referred to as upflows and downflows, respectively, with the words updraughts and downdraughts reserved for more coherent structures.

\subsection{Variation with the Rayleigh number}

\newcommand*{\putabcdef}{
\put(-0.85,1.01)  {(a)}
\put(-0.42,1.01)  {(b)}
\put(-0.85,0.61) {(c)}
\put(-0.42,0.65) {(d)}
\put(-0.85,0.32) {(e)}
\put(-0.42,0.32)   {(f)}
}
\begin{figure}
 \centering
 \includegraphics[width=0.85\textwidth]{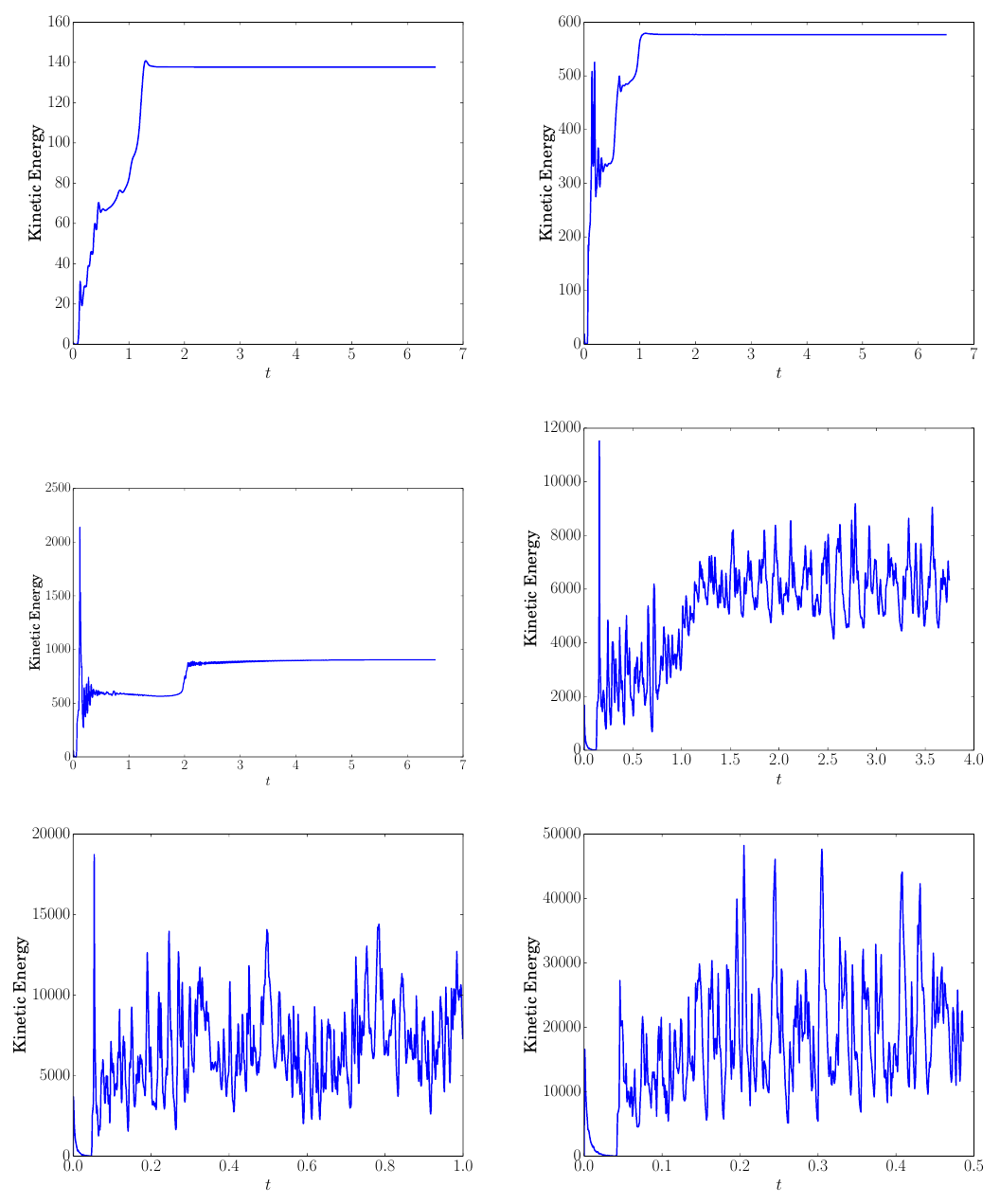}
 \putabcdef
  \caption{Timeseries for nondimensional kinetic energy for numerical simulations with  $\gamma=0.19$ and (a) $Ra=2 \times 10^5$, (b) $Ra=1 \times 10^6$, (c) $Ra=2 \times 10^6$, (d) $Ra=2 \times 10^7$, (e) $Ra=5 \times 10^7$, (f) $Ra=2.5 \times 10^8$. The horizontal axis is time in diffusive units of $H^2/\kappa$, and note the differing scales of the axes in the panels.}
  \label{fig:ke_vs_time}
\end{figure}  
In these experiments we vary the Rayleigh number but keep $\gamma$ fixed, at $\gamma=0.19$.  These parameters give the possibility of continued instability of the drizzle solution to moist convection for sufficiently high Rayleigh numbers, as can be seen from the right-hand panel of \figref{fig:drizz_cont} and \figref{fig:drizzle-stab}.   The initial conditions are typically chosen so that $q=0$ in the interior of the domains except for a parcel of saturated air initially placed near the bottom boundary. This parcel immediately leads to heating and changes in the buoyancy field locally and triggers the start of moist convection. This initial condition naturally gives rise to plume triggering in the quiescent region of the domain before a statistically steady state is achieved, with the subsequent evolution depending on the Rayleigh number.

\renewcommand*{\putabcdef}{
\put(-0.95,1.05)  {(a)}
\put(-0.44,1.05)  {(b)}
\put(-0.95,0.7)  {(h)}
\put(-0.44,0.7) {(d)}
\put(-0.95,0.34) {(e)}
\put(-0.44,0.34)  {(f)}
}
\begin{figure}
 \centering
       \includegraphics[width=\textwidth]{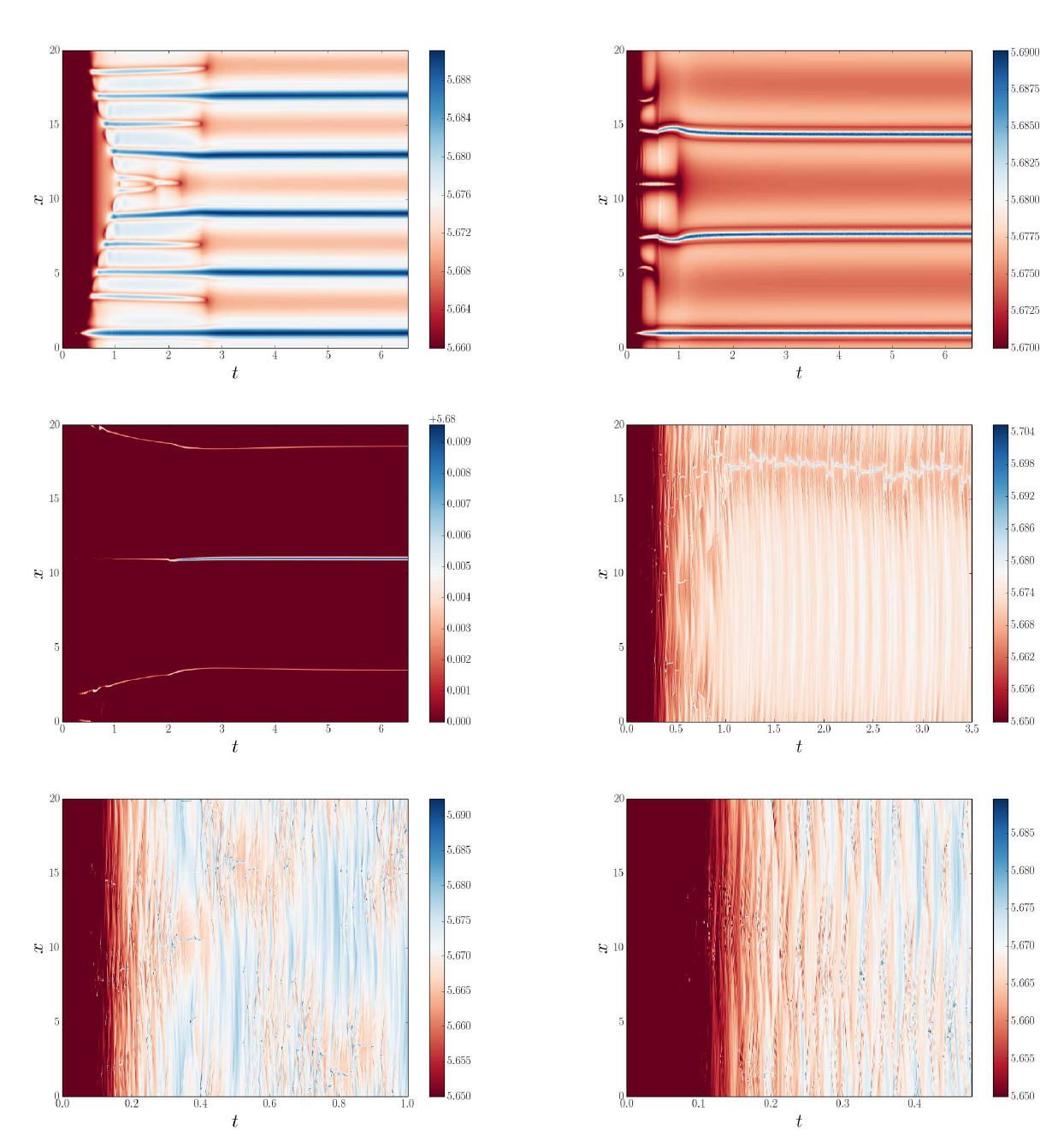}
       \putabcdef
  \caption{As for  Fig.\ \ref{fig:ke_vs_time} but now time series (H\"ovmoller plots) at $z = 0.5$ and for all $x$ for buoyancy  for $\gamma=0.19$ at (a) $Ra=2 \times 10^5$, (b) $Ra=1 \times 10^6$, (c) $Ra=2 \times 10^6$, (d) $Ra=2 \times 10^7$, (e) $Ra=5 \times 10^7$, (f) $Ra=2.5 \times 10^8$.}
  \label{fig:hovm}
 \end{figure}

\subsubsection{Time dependence and basic properties}
For small enough Rayleigh Numbers (e.g., $Ra = 10^2$) the action of diffusion damps out all motions and the system relaxes back to the drizzle solution, with saturation occurring everywhere in the domain. For high enough $Ra$ the drizzle solution is unstable to small perturbations and  the  trigger leads to sustained convection. Figure~\ref{fig:ke_vs_time} gives timeseries of the kinetic energy for increasing $Ra$, and figures \ref{fig:hovm} and  \ref{fig:hovm_RH} show the corresponding buoyancy and relative humidity evolution with time at $z=0.5$ and using H\"ovmoller plots (i.e., space-time plots showing a time series). 
For the three smallest values of $Ra$ the solution reaches a steady state and the kinetic energy approaches a constant value, which increases with $Ra$. For the three higher values of the $Ra$, the solution is unsteady and spatio-temporally modulated and as expected the average kinetic energy increases with $Ra$.  Irregular time-dependence of solutions appears to arise more easily in moist convection than in dry convection, at least in two-dimensions,  because of the sharp nature of the nonlinearities expressed by the thresholds in moist convection.

\renewcommand*{\putabcdef}{
\put(-0.84,0.97)  {(a)}
\put(-0.42,0.97)  {(b)}
\put(-0.84,0.64)  {(c)}
\put(-0.42,0.64) {(d)}
\put(-0.84,0.31) {(e)}
\put(-0.42,0.31)   {(f)}
}

\begin{figure}
 \centering
      \includegraphics[width=0.85\textwidth]{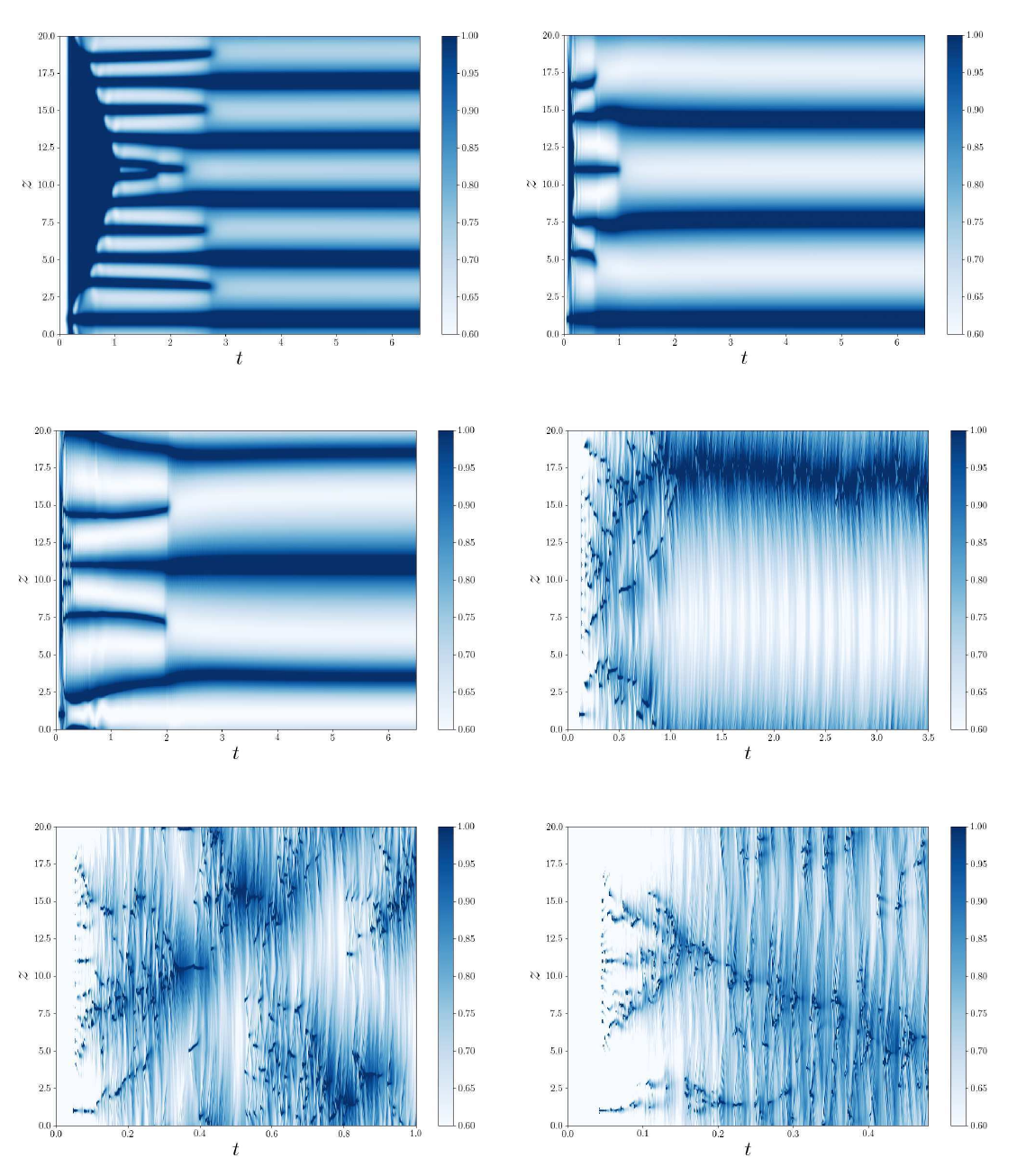}
       \putabcdef
  \caption{Time series (H\"ovmoller  plots) for relative humidity ($q/q_s$) at $z=0.5$ and for all $x$ for $\gamma=0.19$ at (a) $Ra=2 \times 10^5$, (b) $Ra=1 \times 10^6$, (c) $Ra=2 \times 10^6$, (d) $Ra=2 \times 10^7$, (e) $Ra=5 \times 10^7$, (f) $Ra=2.5 \times 10^8$.}
  \label{fig:hovm_RH}
 \end{figure}

The nature of time-dependence in the solutions can be seen in some of the movies that are included in the supplementary material.  Movie 1 shows the evolution from the initial conditions for the four variables, $b$, $q$, $T$ and $u$, for $Ra = 2 \times 10^5$. Initially the moist blob is seen in the $q$ field and rapidly heats the parcel causing a deficit in the buoyancy field that leads to its rise. This buoyant rise triggers gravity waves that advect the moisture field and thus triggers further buoyant patches to rise in the vicinity of the initial plume. This process continues until several (of the order of ten) buoyant plumes are found in the box. These plumes continue their evolution by sometimes merging and sometimes dying, until a steady state is reached with 5 equally spaced plumes in the box. The evolution is illustrated by  H\"ovmoller plot of the buoyancy $b$ as a function of $x$ and $t$ for fixed $z=0.5$, shown in Figure~\ref{fig:hovm}a. The diagram clearly shows the triggering  of plumes at the beginning of the calculation followed by their merging and approach to a steady state. Figures~\ref{fig:hovm} b,c show that the dynamics is similar for the next two smallest values of $Ra$, but for $Ra= 2 \times 10^6$ there are only 3 plumes in the computational domain at the end of the calculation, with a wider spacing between plumes.  

Movies 2-4 show the evolution for $Ra=2 \times 10^7$, $5 \times 10^7$ and $2.5 \times 10^8$ respectively. They show that the buoyant parcel rapidly triggers turbulent moist convection for these Rayleigh numbers. Here the plumes are much more time-dependent and  for $Ra=2 \times 10^7$ eventually the system is left in a state with a single, extremely time-dependent plume in the computational domain. This plume is very efficient at driving gravity waves, though for these parameters the waves are not instrumental in triggering plumes at other locations in the computational domain (as shown in Figure~\ref{fig:hovm}d). However as the Rayleigh number is increased multiple plumes are found in the domains, triggered by the gravity waves. These plumes are born and die rapidly as shown in the  H\"ovmoller plots of Figures~\ref{fig:hovm}(e,f). The triggering by the gravity waves for the turbulent solutions is most clearly seen in Figure~\ref{fig:hovm_RH}(d-f).

\renewcommand*{\putabcdef}{
\put(-0.97,0.77)  {(a)}
\put(-0.47,0.77)  {(b)}
\put(-0.97,0.51)  {(c)}
\put(-0.47,0.51) {(d)}
\put(-0.97,0.25) {(e)}
\put(-0.47,0.25)   {(f)}
}
 \begin{figure}
 \centering
  \includegraphics[width=\textwidth]{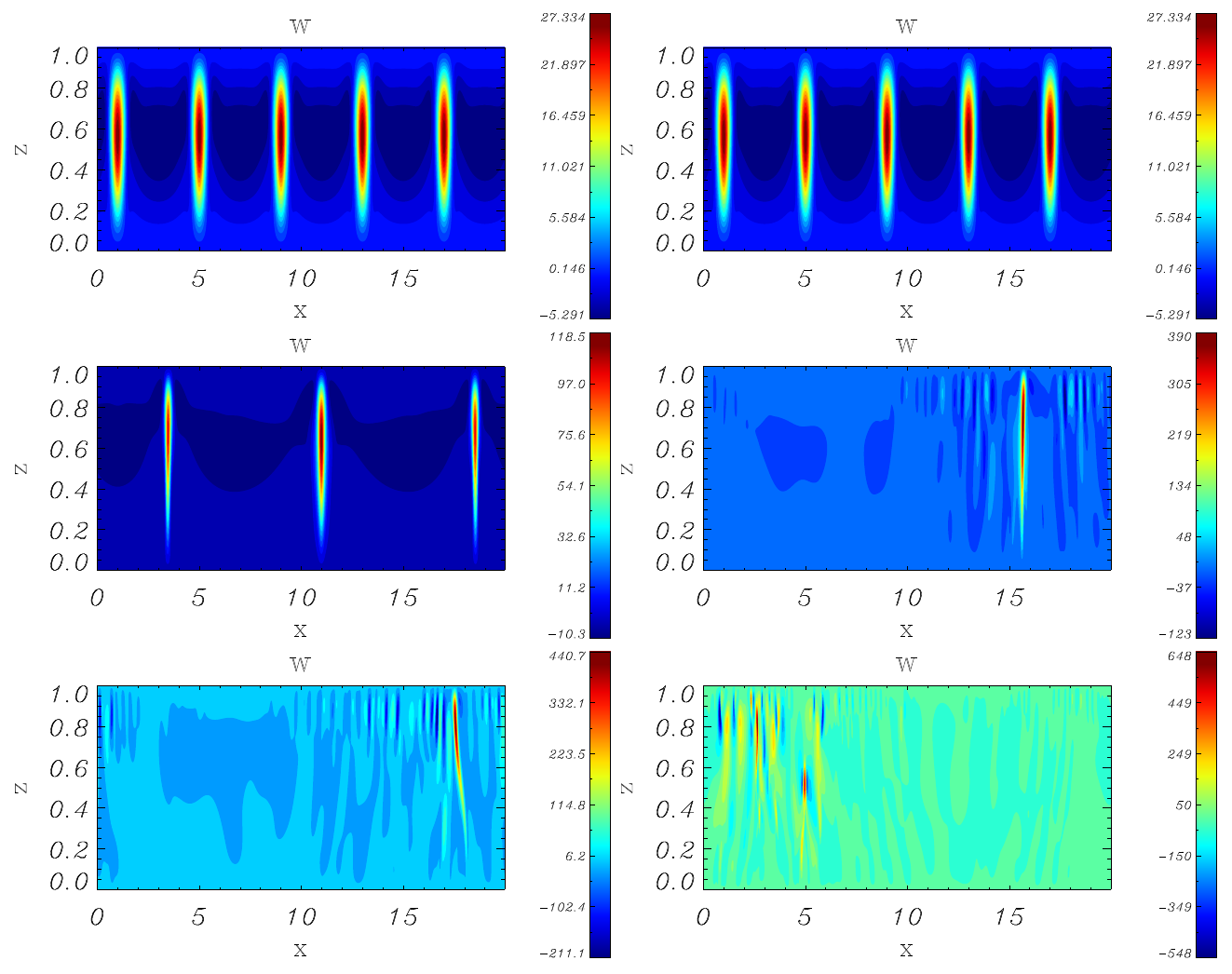} 
  \putabcdef
\caption{Snapshots of vertical velocity in a statistically equilibrated state for $\gamma=0.19$ at (a) $Ra=2 \times 10^5$, (b) $Ra=1 \times 10^6$, (c) $Ra=2 \times 10^6$, (d) $Ra=2 \times 10^7$, (e) $Ra=5 \times 10^7$, (f) $Ra=2.5 \times 10^8$.  } 
 \label{fig:w_final}
 \end{figure}

\renewcommand*{\putabcdef}{
\put(-0.89,1.02)  {(a)}
\put(-0.44,1.02)  {(b)}
\put(-0.89,0.67)  {(c)}
\put(-0.44,0.67) {(d)}
\put(-0.89,0.32) {(e)}
\put(-0.44,0.32)  {(f)}
}
 \begin{figure}
 \centering
     \includegraphics[width=0.9\textwidth]{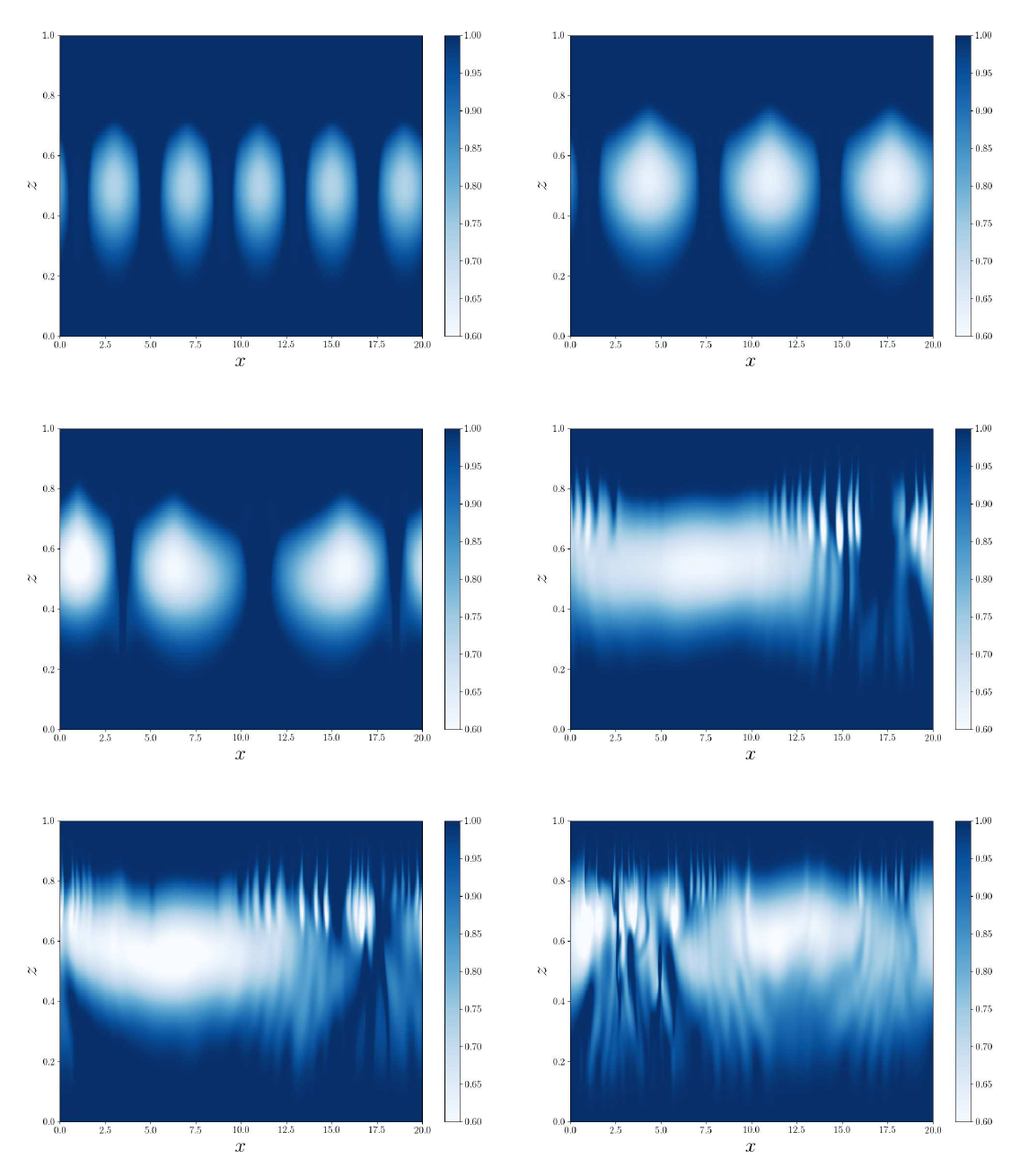}
       \putabcdef
  \caption{Snapshots of relative humidity in a statistically equilibrated state for $\gamma=0.19$ at (a) $Ra=2 \times 10^5$,  (b) $Ra=1 \times 10^6$, (c) $Ra=2 \times 10^6$, (d) $Ra=2 \times 10^7$, (e) $Ra=5 \times 10^7$, (f) $Ra=2.5 \times 10^8$.  }
  \label{fig:rh_final}
 \end{figure}

 Figures~\ref{fig:w_final} and \ref{fig:rh_final} show snapshots of the spatial dependence of the solutions in a statistically equilibrated state for the vertical velocity ($w$) and the relative humidity (defined as $q/q_s(T)$). For the steady solutions the solution takes the form of strong narrow updraughts surrounded by broader gentler downdraughts. The asymmetry is well-known result in moist atmospheric convection \citep[e.g.,][]{Ludlam80}, but in Boussinesq \RB convection the updraughts and downdraughts have exactly the same properties because of the up-down symmetry in the problem. 
 
 As the Rayleigh number is increased the updraughts become thinner and sparser and at high enough Rayleigh numbers they become unstable. For these higher Rayleigh numbers the plumes are extremely thin (and as noted earlier, very time-dependent). The relative humidity shows that a moderate Rayleigh number the flow is saturated (red colour)  in the updraughts and in the  diffusive layers at the top and bottom of the computational domain. The downdraughts are in general not saturated since they are bringing dry air down into regions with higher temperatures where $q_s$ is larger, and the horizontally averaged relative humidity is a minimum in the domain interior, as discussed more below.
  
  \newcommand*{\putab}{
  \put(-0.98,0.44) {(a)}
  \put(-0.49,0.44) {(b)}
  }
 \begin{figure}
 \centering
 \includegraphics[width=0.49\textwidth]{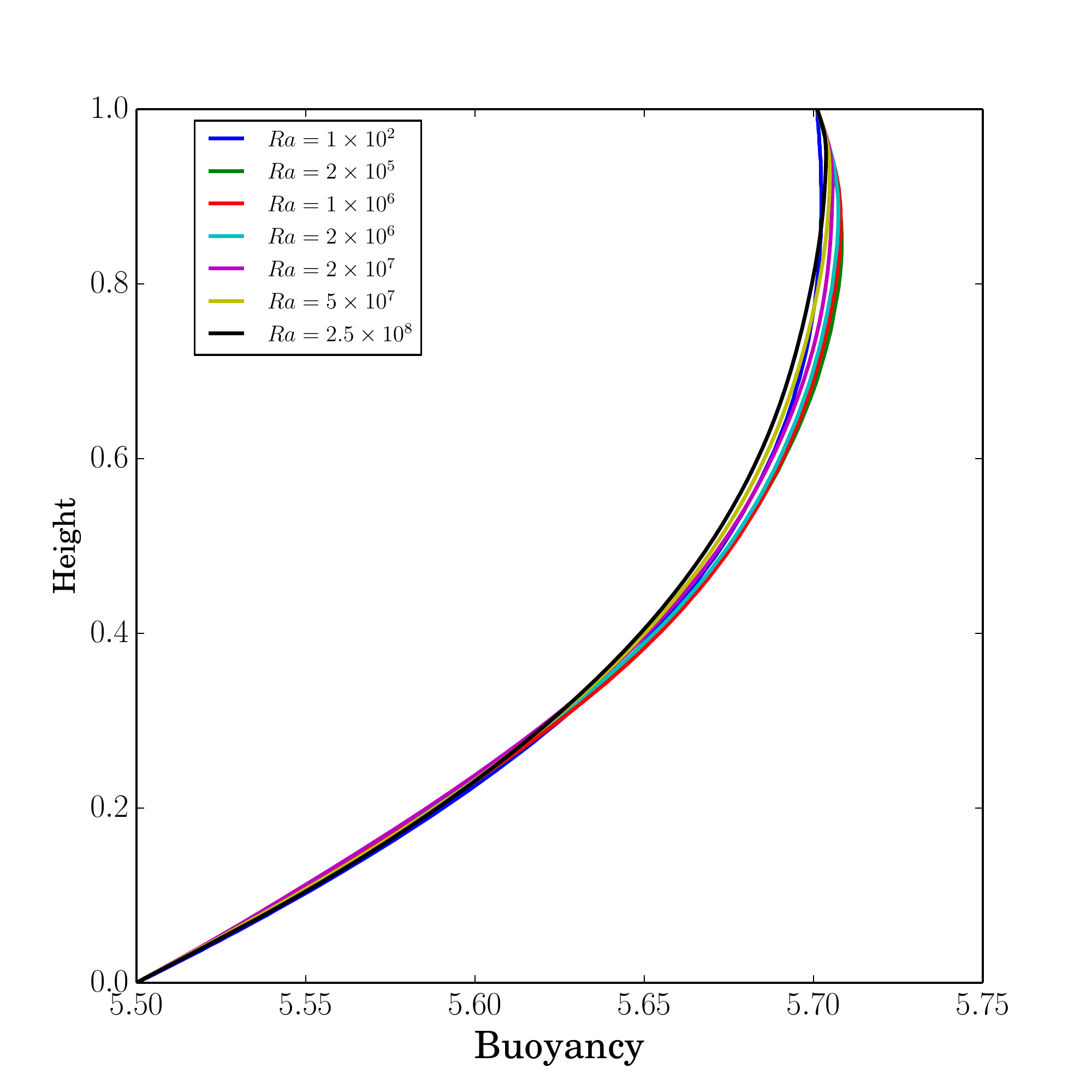}
 \includegraphics[width=0.49\textwidth]{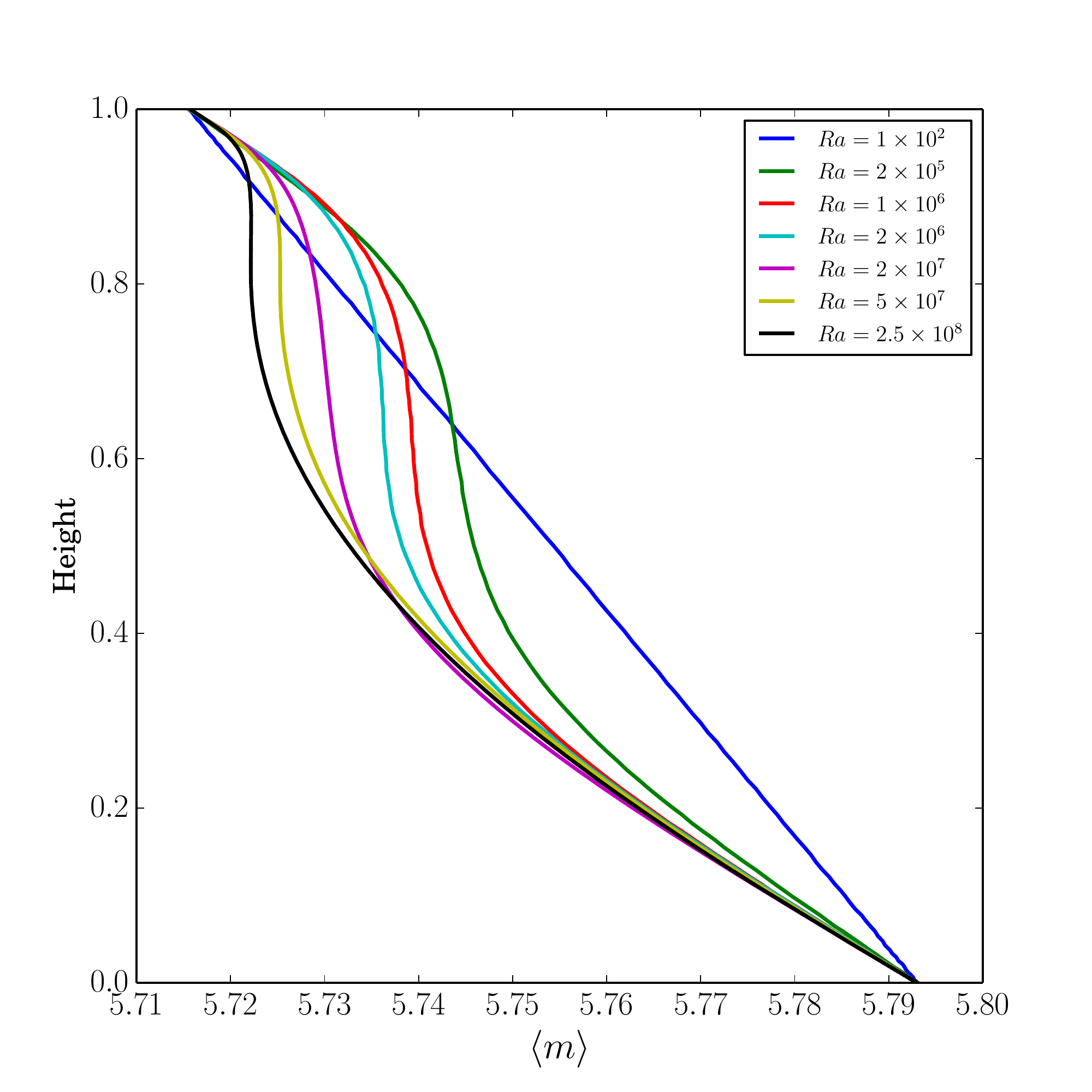}
  \putab
  \caption{Horizontally averaged profiles for (a) buoyancy and (b) $m(z)$ for $\gamma=0.19$ and a range of $Ra$.}
  \label{fig:mean_prof}
 \end{figure}

\subsubsection{Equilibrated properties}

Equilibration of the moist convective solution is more complicated than in the dry problem. Roughly speaking, dry \RB convection  equilibrates by mixing the buoyancy to an adiabatic profile in the core, thus effectively removing the driving except in thermal boundary layers at the top and bottom of the domain leaving the neutral profile,  $\ppp bz \approx 0$, in the interior. In moist convection the analogous solution is a state corresponding to the saturated adiabatic lapse rate.  In atmospheric models that cannot resolve convection, a simple convective parametrization is to adjust the buoyancy profile to satisfy these criteria, although adjustment methods are now commonly regarded as over-simple because the implied separation of timescales is too extreme (although improved parameterizations are hard to come by). To get a sense of the equilibration processes here we will show how various profiles vary as a function of Rayleigh number, with detailed analysis left for future work.

\begin{figure}
    \centering
    \includegraphics[width=0.5\textwidth]{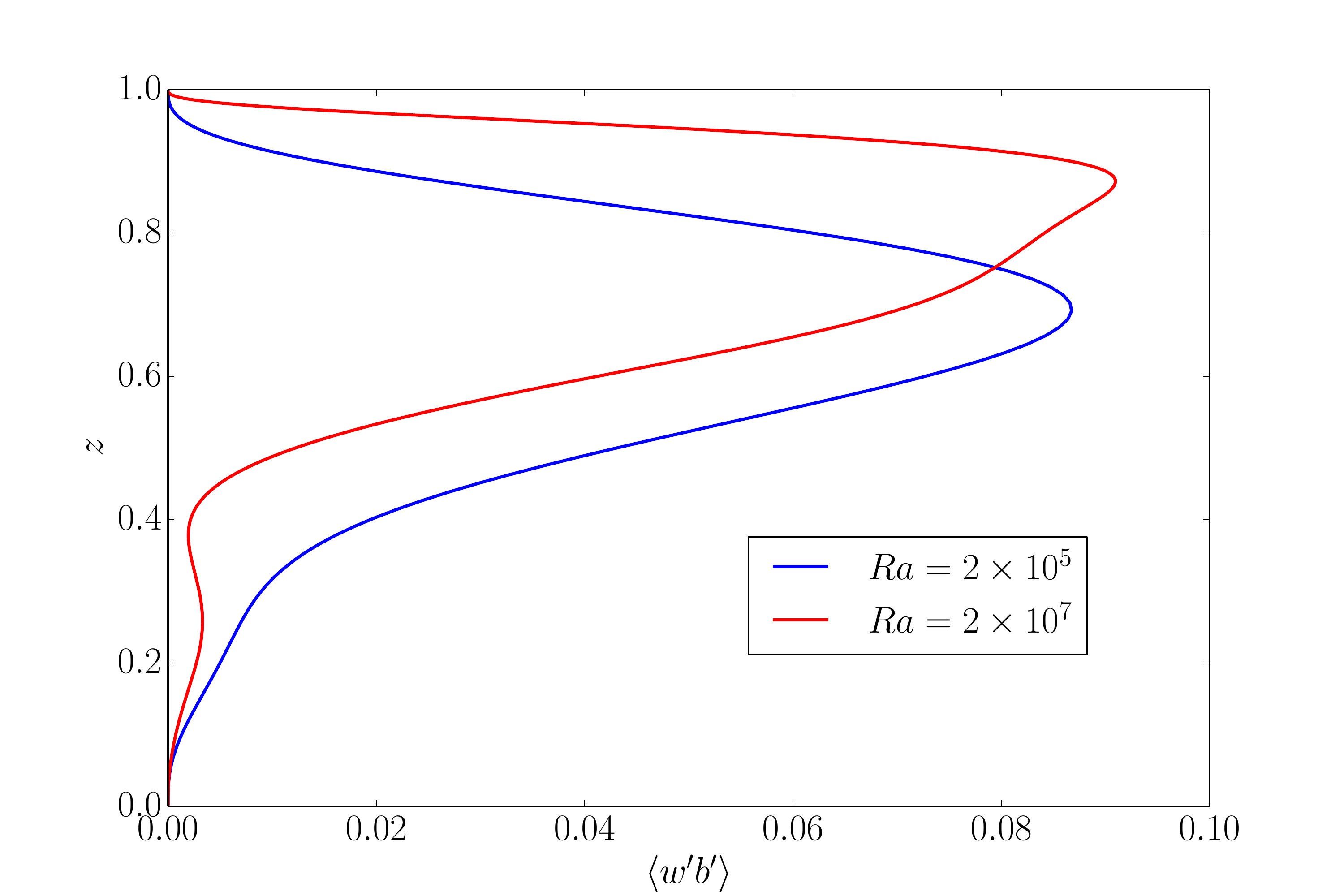}
    \caption{Time and horizontally-averaged vertical buoyancy flux, $\avg{w'b'}$, at two values of Rayleigh number. Note the peak flux rising up to the top boundary layer at higher Rayleigh number. The time series of these fields are shown in \figref{fig:bwhoff}.}
    \label{fig:bwave}
\end{figure}

\begin{figure}
    \centering
 \includegraphics[width=\textwidth]{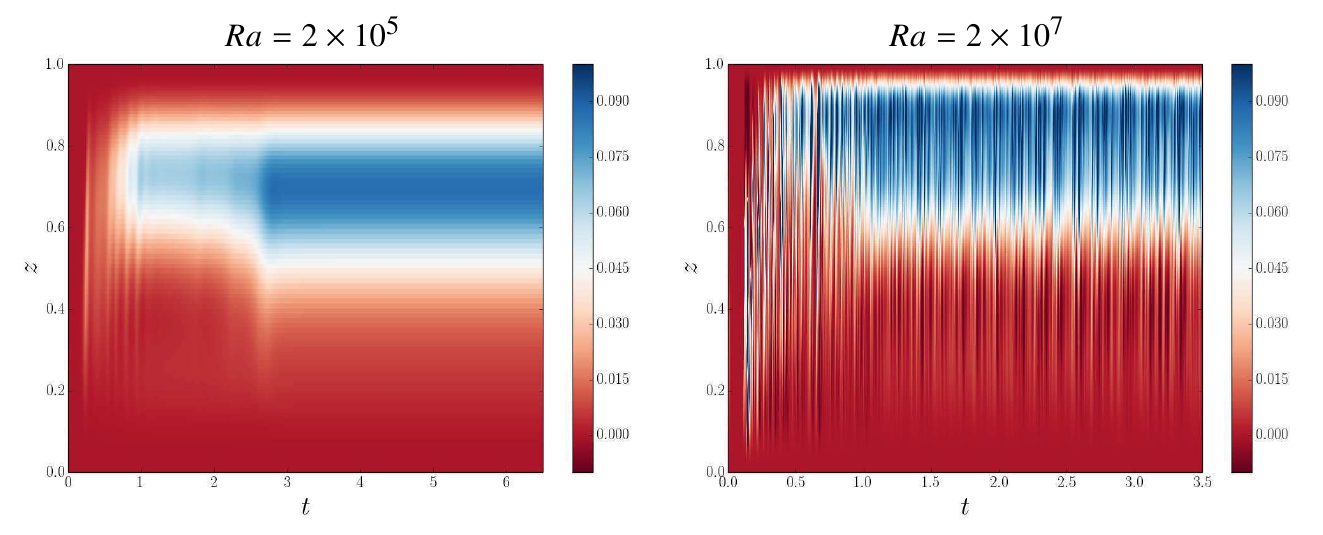}
    \caption{Time series (H\"ovmoller plots) of the $x$-averaged buoyancy fluxes $\avg{w'b'}$ at each height,  at two values of Rayleigh number. The time-averaged fields, over the later parts of the time period, are shown in \figref{fig:bwave}.}
    \label{fig:bwhoff}
\end{figure}

Figure~\ref{fig:mean_prof} shows the horizontally averaged profiles in the statistically equilibrated state for the buoyancy and $m$.  Both are significantly modified by convection --- recall that $m$ has a linear profile in $z$ in the  saturated drizzle solution, which is reproduced only by the $\Ra = 1\times 10^2$ solution.  The corresponding profiles in the upflows and downflows alone are shown in \figref{fig:mean_updown}.  These profiles show fairly uniform values of $m$ in the upper part of the domain, but the degree to which this is caused by mixing and/or adjustment  through gravity waves in subsiding regions will be investigated in a subsequent paper. The lower part of the domain remains unstable even at high Rayleigh number (see also \figref{fig:mean_prof_M}c), and whether this result holds at still higher Rayleigh number, and in three dimensions, is also a topic for future investigation.  The buoyancy fluxes giving rise to these profiles are shown in \figref{fig:bwave}  and \figref{fig:bwhoff}. In both the steady and unsteady cases the buoyancy flux is peaked in the upper half of the domain, in fact close the diffusive upper boundary layer in the unsteady case. The moisture flux (not shown) peaks lower down, largely because there is much more moisture where the temperature is higher.     

 \newcommand*{\putabcd}{
 \put(-0.98,1.01) {(a)}
 \put(-0.49,1.01) {(b)}
 \put(-0.98,0.51) {(c)}
 \put(-0.49,0.51) {(d)}
  }
 \begin{figure}
 \centering
 \includegraphics[width=0.99\textwidth]{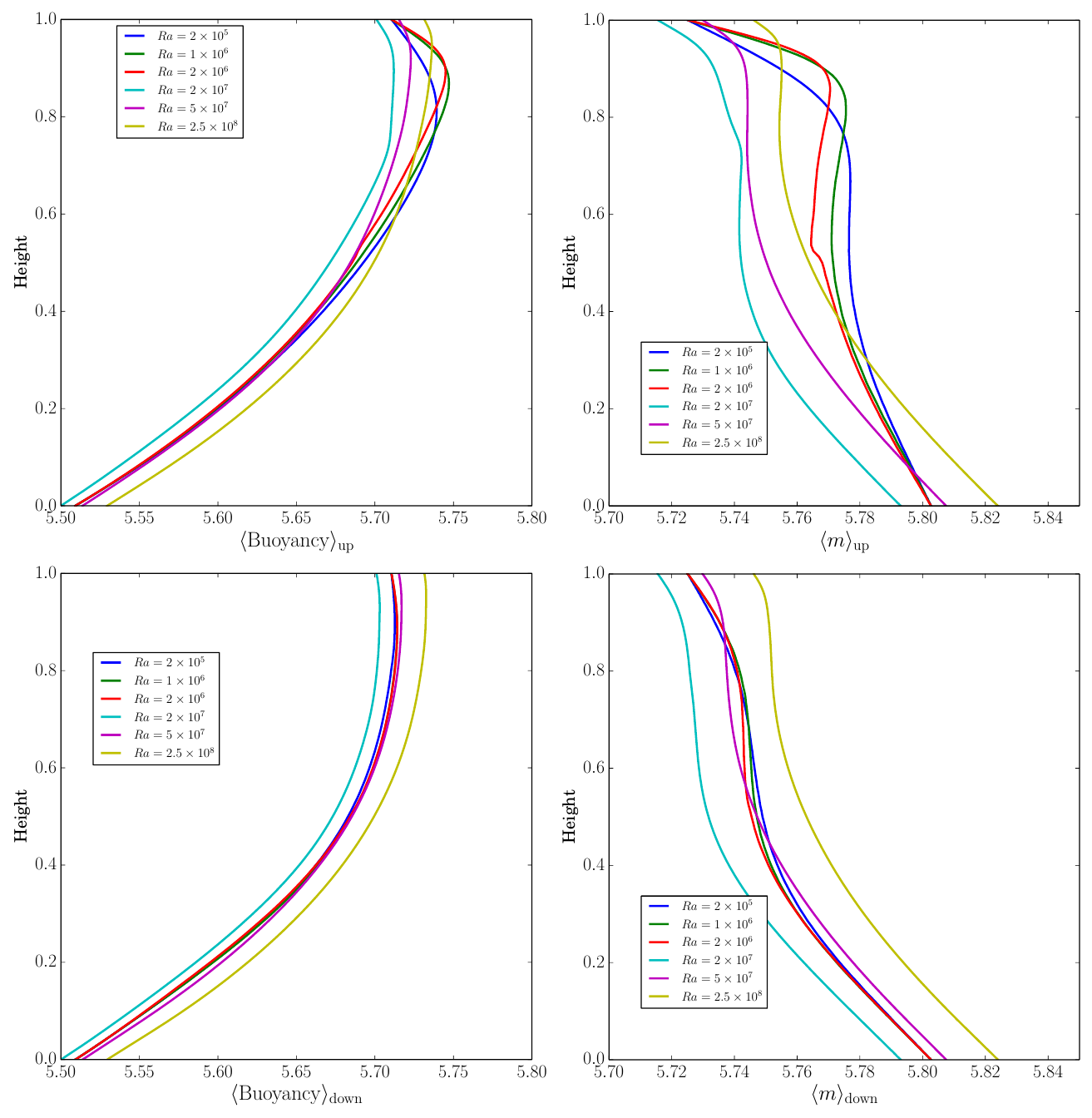}
 \putabcd
 \caption{Top row: Averages taken over the upflows alone (where $w > 0$) of (a) the buoyancy and (b) the moist static energy, $m$.  Bottom row: Same, except that averages are taken over downflows alone (where $w <  0$).}
  \label{fig:mean_updown}
 \end{figure} 
 
 \renewcommand*{\putab}{
 \put(-1.,0.34) {(a)}
 \put(-0.49,0.3) {(b)}
  }
\begin{figure}
 \centering
 \includegraphics[width=0.49\textwidth]{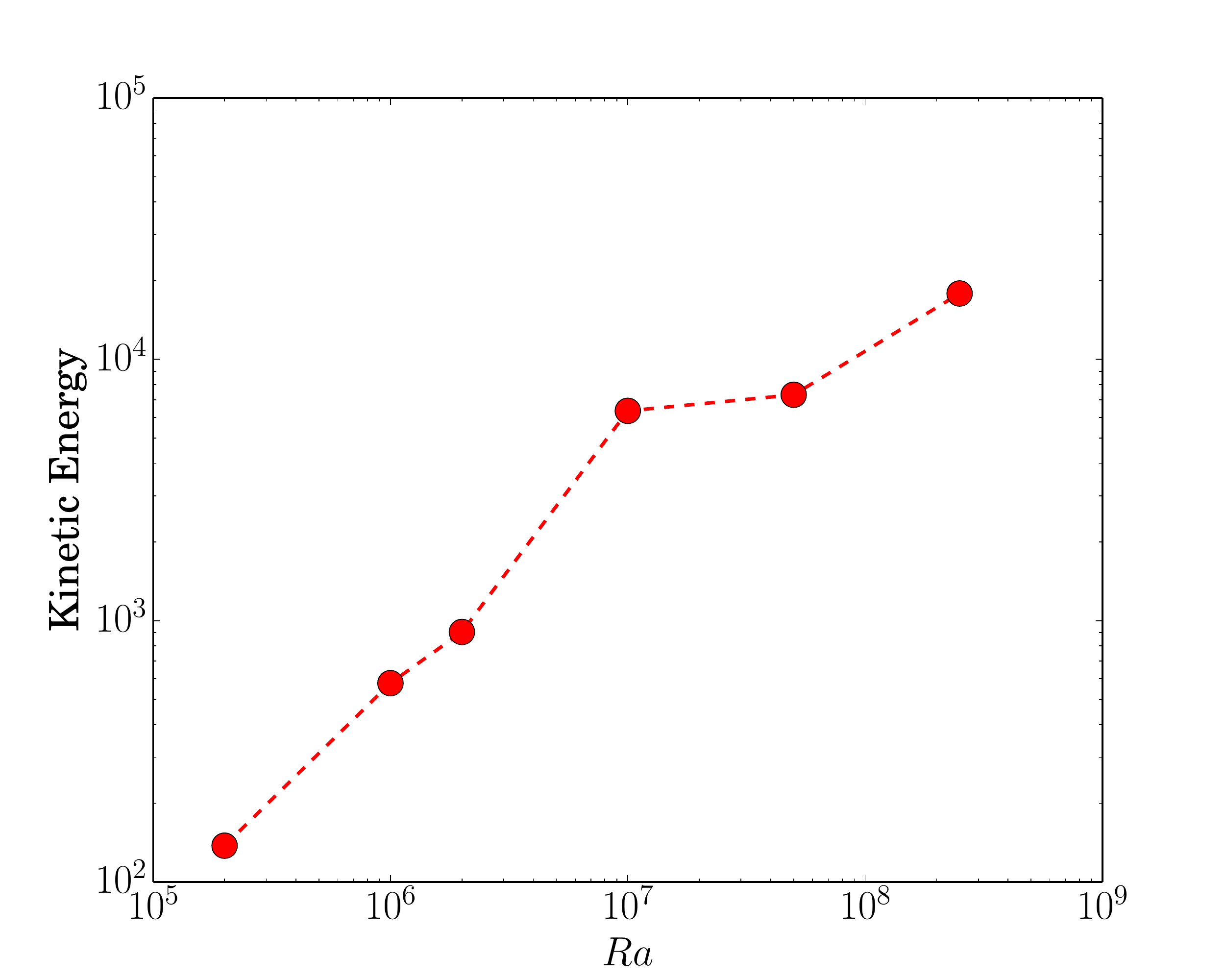}
 \includegraphics[width=0.49\textwidth]{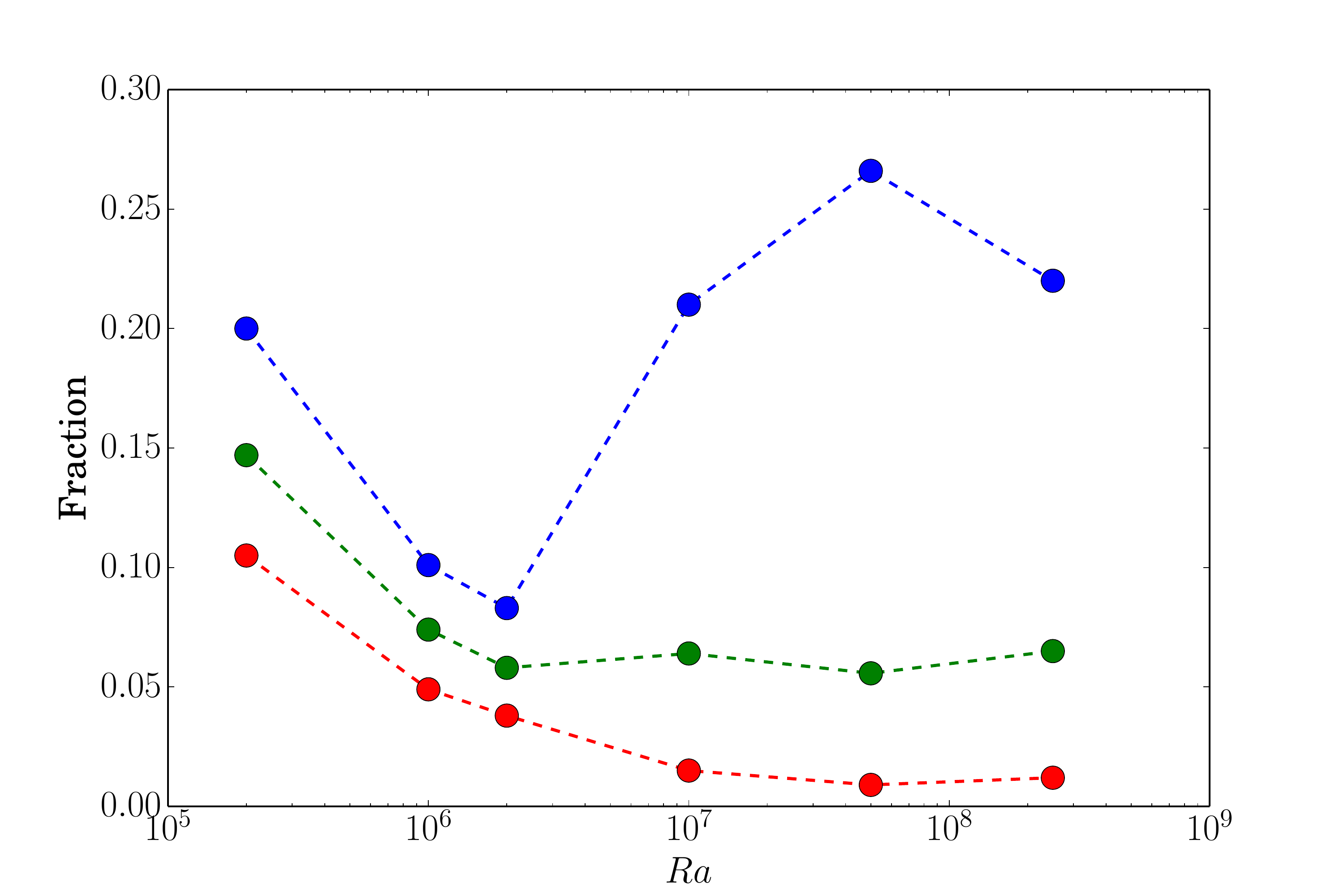}
 \putab
  \caption{(a) Kinetic energy as a function of $Ra$. (b) Area fraction of near-saturated upflows as a function of $Ra$, with three different criteria. The blue line requires $w > 0$ and $RH >0.95$, the green line requires $w > 0.02 w_\text{max}$ and $RH > 0.95$ and red line requires $w > 0.1 w_\text{max}$ and $RH > 0.95$.
  }
  \label{fig:kefrac}
 \end{figure}

 Figure~\ref{fig:kefrac} shows the kinetic energy and the fraction of the domain covered by near saturated upflows as a function of $Ra$. As expected (at least for moderate values of Rayleigh number, and given that the Reynolds number is often taken to be the square root of the Rayleigh number) the kinetic energy increases with $Ra$.   The area fraction of upflows has a more complicated dependence, and depends on the criterion used.  For the steady solutions (the lowest three values of $Ra$) the area fraction actually decreases with $Ra$ if the weakest criterion ($w>0$) is used.  As $Ra$ is increased further the solutions become time-dependent and the area fraction of upflows stays roughly constant --- the strong upflows are very narrow, but there are a greater number of weak upflows. However, this criterion is most likely picking up flows with positive vertical velocity due to gravity waves and not associated with updraughts, and this V-shaped dependence is in fact reproduced without any restriction on relative humidity.   With  the more restrictive criteria (red and green lines in the figure) the upflow fraction decreases and then seems to converge to a definite value as Rayleigh number increases, with the structures generally remaining space filling.  Understanding the parameter dependence of these results, and  whether there is any universality, is of direct relevance to meteorology, since `mass flux parameterization schemes'  commonly used in coarse resolution atmospheric models often assume that updraughts occupy some fixed fraction of the domain.  

 \begin{figure}
     \centering
     \includegraphics[width=0.49\textwidth]{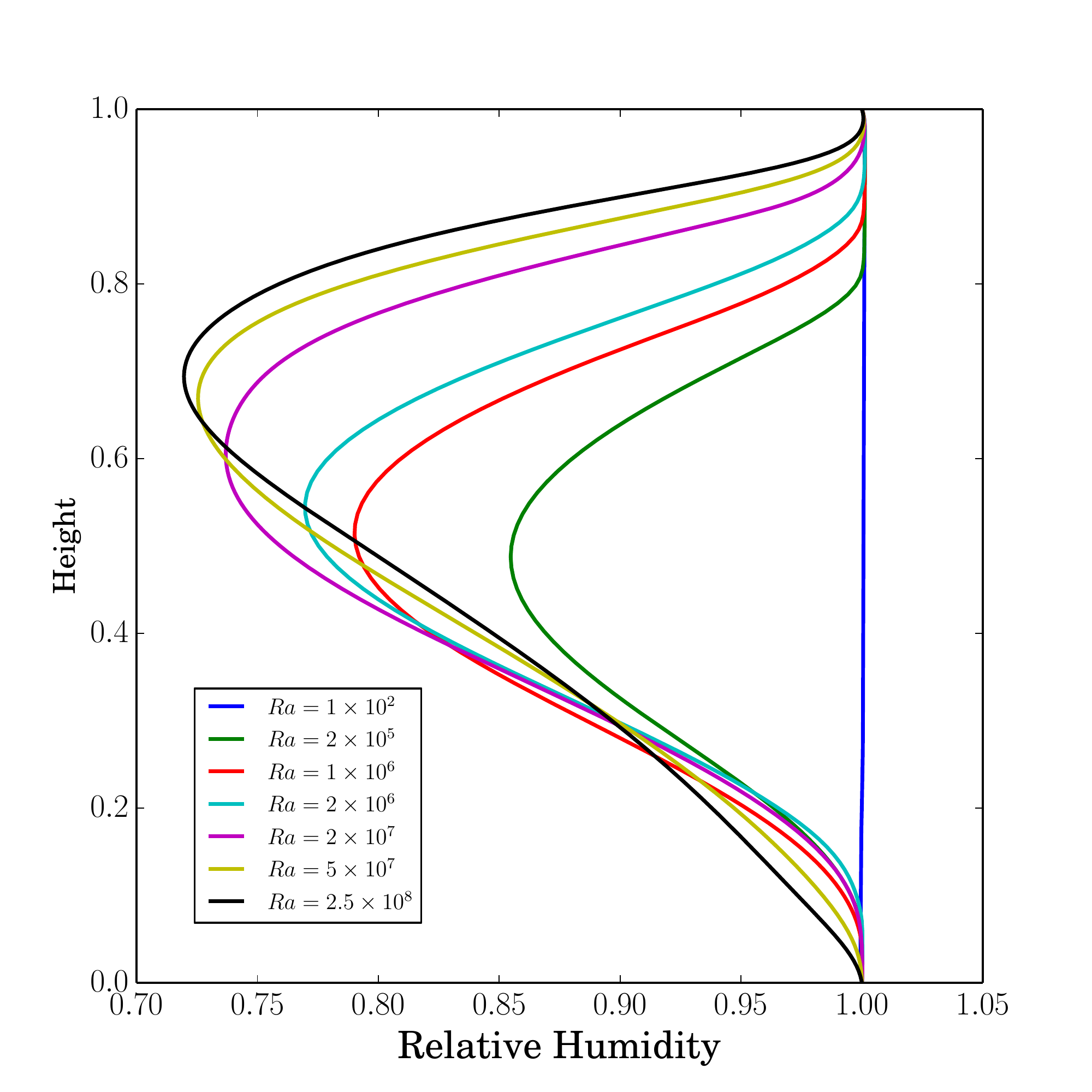}
     \caption{ Horizontally averaged relative humidity over a range of Rayleigh numbers, with all other parameters fixed. }
     \label{fig:mean-rh}
 \end{figure}

\newcommand*{\putRa}{
\put(-0.66,0.87) {{$Ra=2 \times 10^5$}}
\put(-0.18,0.87) {{$Ra=1 \times 10^6$} }
\put(-0.66,0.47) {{$Ra=2 \times 10^6$}}
\put(-0.18,0.47) {{$Ra=2 \times 10^7$}}
\put(-0.66,0.07) {{$Ra=5 \times 10^7$}}
\put(-0.2,0.07) {{$Ra=2.5 \times 10^8$}}
}
\begin{figure}
    \centering
 \includegraphics[width=0.95\textwidth]{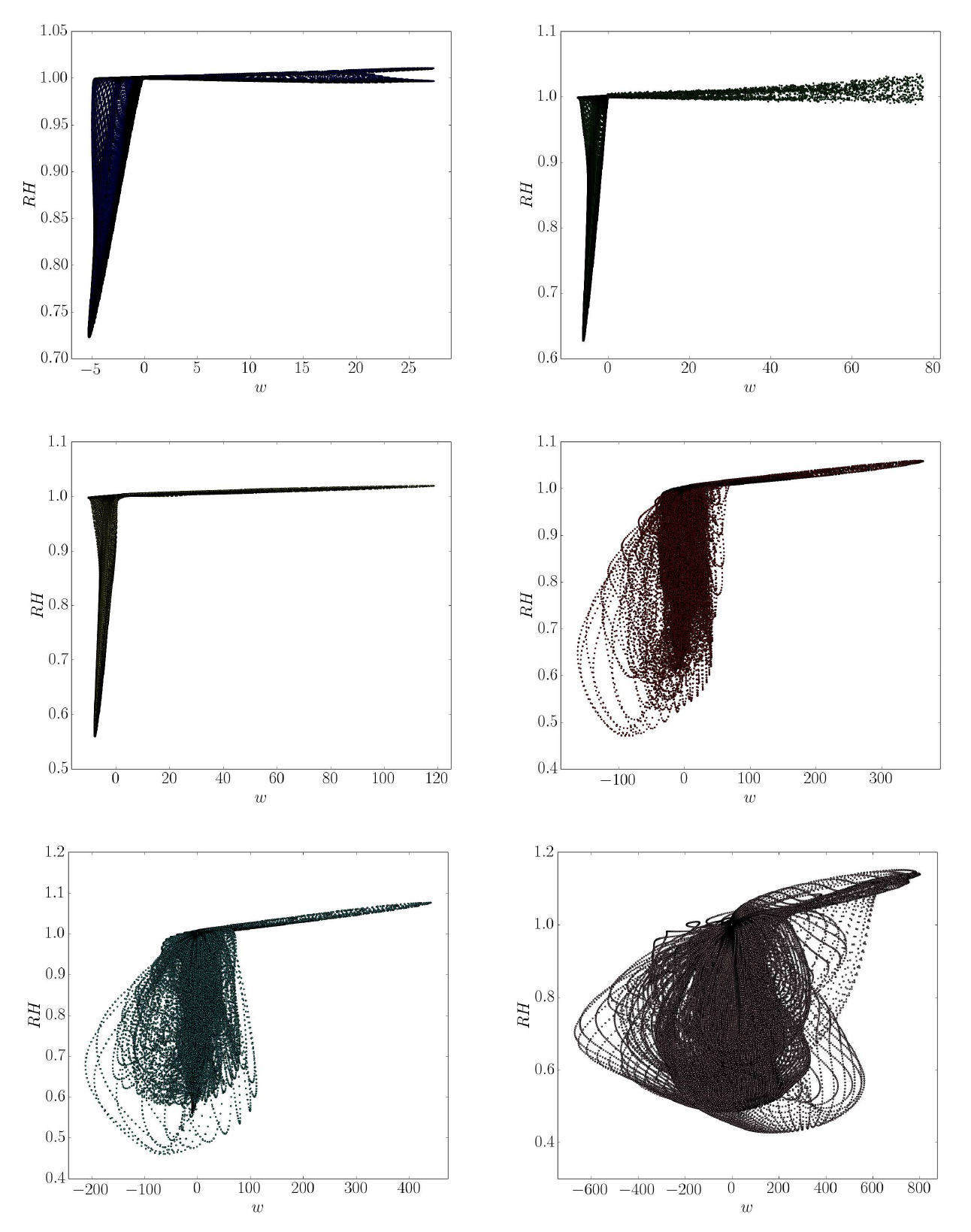}
\putRa
     \caption{Scatter plots of relative humidity, $RH$,  and vertical velocity, $w$, over a range of Rayleigh numbers, as labelled. Note the changes in the scale of the $x$-axis.}
     \label{fig:scatter}
\end{figure}

\subsubsection{Relative humidity}

Relative humidity has no analogue whatsoever in the dry system but is of fundamental importance to the climate system.  In the fast condensation limit, and at high Reynolds number,  the relative humidity of a parcel is largely determined by the temperature of the location at which it was last saturated, since in the absence of condensation or re-evaporation (which in some situations is quite important) a parcel conserves its specific humidity.  Thus, updraughts tend to be saturated (the parcels are moving to lower temperatures) and downdraughts unsaturated, but quantitatively determining the vertical profile is non-trivial, especially as the flow becomes turbulent and the updraughts entrain dryer air from the surroundings. 

In our system, there is a relative humidity minimum at mid-levels and the values in the minimum decrease, albeit slowly, as  Rayleigh number is increased (\figref{fig:mean-rh}). Such a minimum is also seen in atmospheric observations in convecting regions and is discussed by \citet{Romps14}. This asymmetric C-shaped profile arises in our simulations for two reasons. The first is that the fraction of the domain covered by upflows decreases as the Rayleigh number increases \figref{fig:kefrac}. For the laminar flows the upflows remain saturated  so that a decrease in their volume fraction is significant. For the unsteady cases the average relative humidity of the upflows tends to decrease with Rayleigh number,  a signature of the turbulent vertical upflows entraining relatively dry material into their flanks from the ambient background, even as the core remains saturated..  This entrainment of relatively dry air also contributes to the decrease of $m$ in the upflows. The second factor is more specific to this model --- as the Rayleigh number increases,  the turbulent convection mixes moisture more effectively and the boundary layer at the top of the domain where the atmosphere is forced to return to a saturated state by the top boundary conditions becomes thinner.  However, numerical simulations with an upper boundary condition of no moisture flux (i.e., $\ppp q z= 0$ at $z=1$) in fact show similar behaviour, in particular having a similar C-shaped relative humidity profile with minimum in the domain interior (not shown), so this effect may be less important. 

Variations in the relative humidity distribution may be graphically illustrated with a series of scatter-plots of $w$ and $RH$ at various values of $Ra$ (\figref{fig:scatter}). At the lower values of $Ra$ the distributions exhibit two narrow limbs: the strongly-ascending air is close to saturated (top, near horizontal limb), while the air that  is significantly sub-saturated is all weakly subsiding (left, near vertical limb). In this regime, as $Ra$ increases from $Ra=2 \times 10^5$ to $Ra=2 \times 10^6$, the updraughts get stronger and the regions of subsidence get drier. At $Ra=2 \times 10^7$ the distribution of $w$ and $RH$ changes significantly with the onset of unsteadiness leads to the generation of gravity waves in the domain (c.f., \figref{fig:ke_vs_time}d). In the subsaturated limb there are now wide distributions of $w$, representing both mean subsidence and regions of local ascent, both with broad relative humidity distributions. In this unsaturated part of the domain $RH$ increases with upward parcel displacement and thus for a gravity-wave--like solution ---  where $w$ and displacement are out of phase --- the solution forms closed loops. As $Ra$ increases further, in panels (d) to (f) of \figref{fig:scatter},  the solutions become increasingly turbulent and the maximum $w$ and minimum $RH$ in the two limbs of the distribution become more extreme (note the scales of the axes are changing). Interestingly, for the very highest $Ra$ there are points on the margins of the cloudy updraughts that appear to transition smoothly from conditions of vigorous, saturated ascent (high $w$ and $RH\approx 1$) to the ambient conditions of lower $RH$ and $w$. In this regime, the cloudy updraughts are becoming diffused, so that they are carrying subsaturated air on their flanks, a feature that is not seen at the lower values of $Ra$.  The updraughts in this turbulent regime are entraining dryer, unsaturated air into their margins, giving rise to a the broader distribution of relative humidity in their unsteady flanks, but the presence of gravity waves is still seen in the distinctive patterns of the scatter plots.

\subsection{Variation with condensational parameter, $\gamma$}

 \newcommand*{\putabc}{
 \put(-1.,0.25) {(a)}
 \put(-0.66,0.25) {(b)}
 \put(-0.33,0.25) {(c)}
  }
  \begin{figure}
 \centering
   \includegraphics[width=0.33\textwidth]{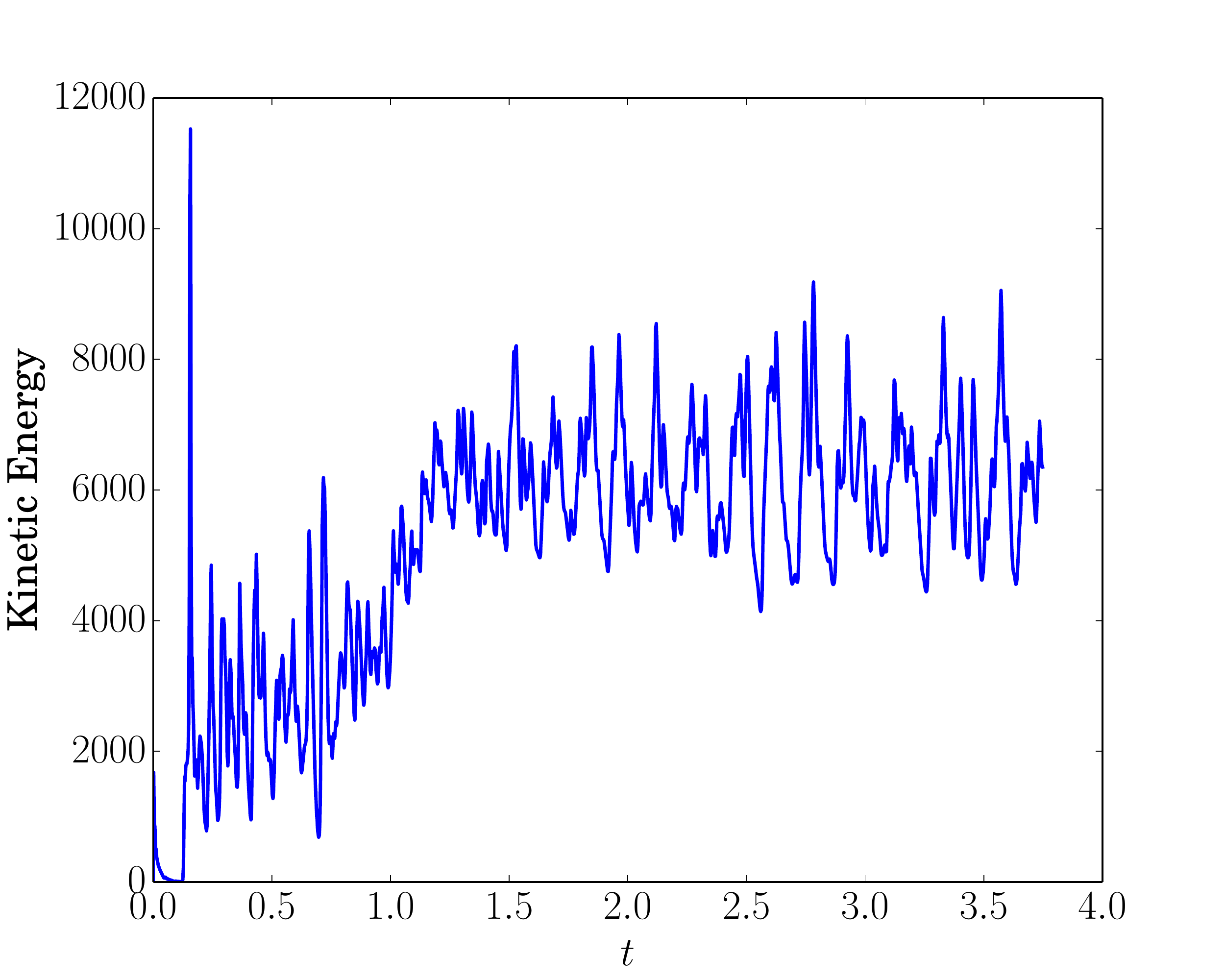}%
   \includegraphics[width=0.33\textwidth]{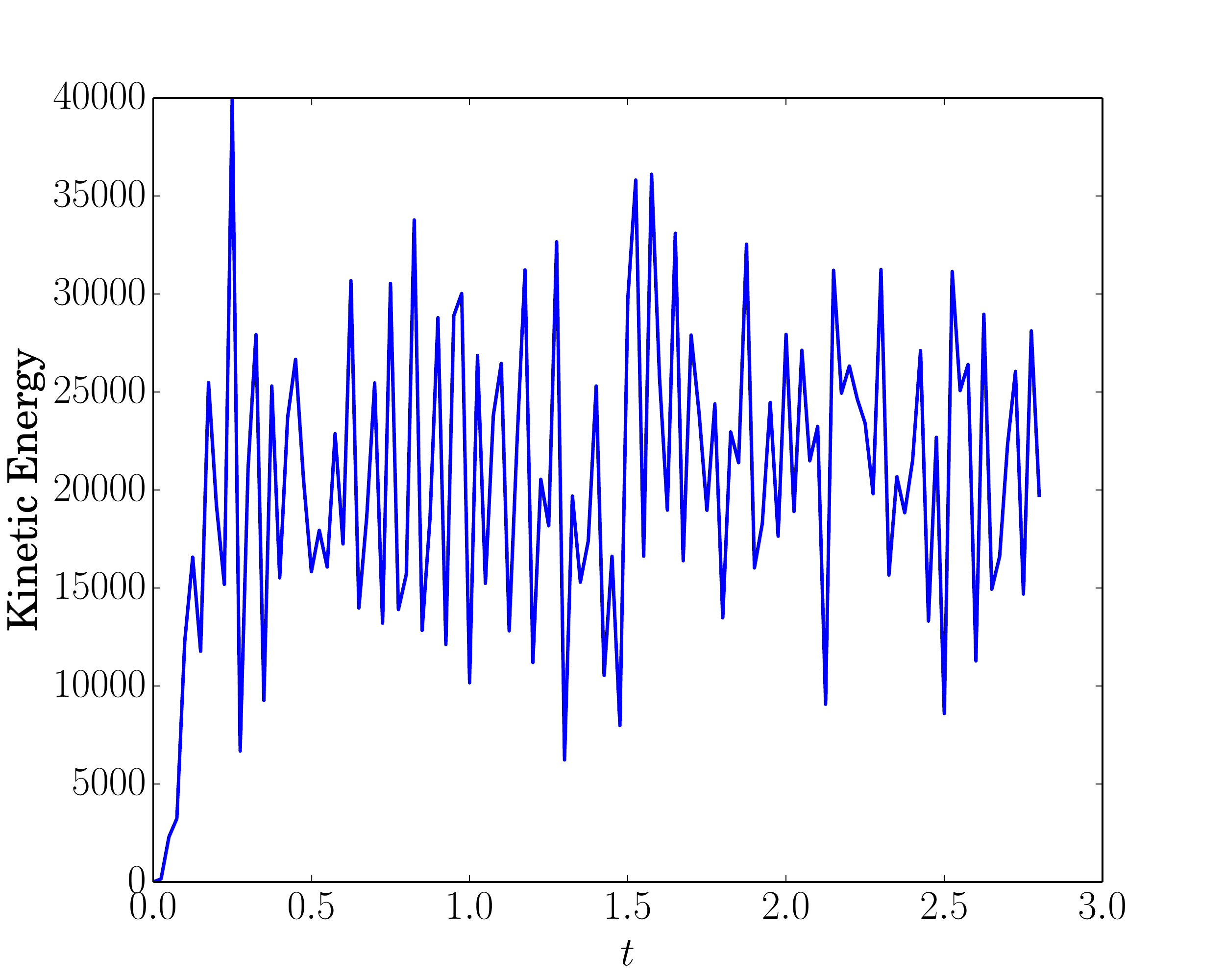}%
   \includegraphics[width=0.33\textwidth]{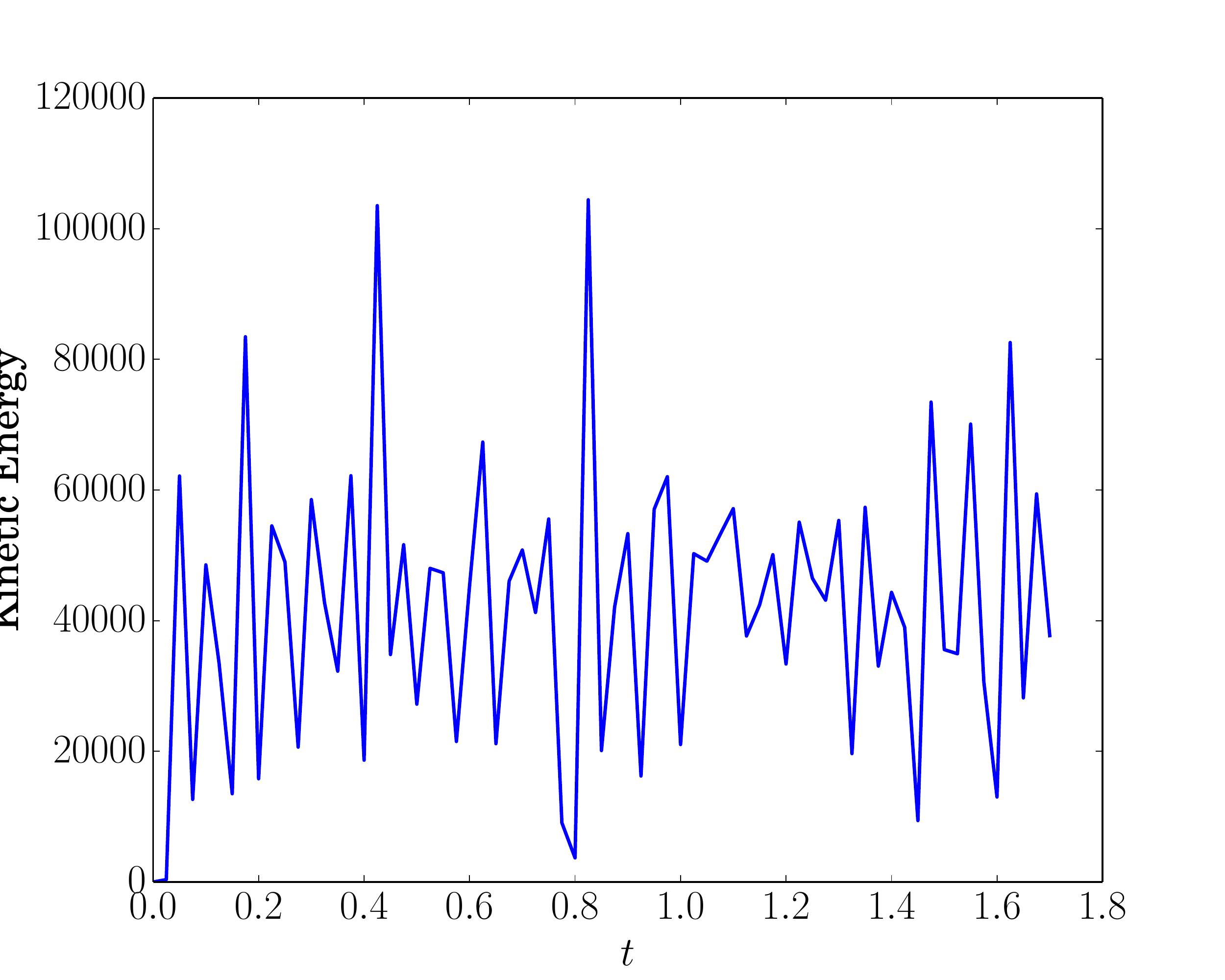}%
\putabc
  \caption{Timeseries for kinetic energy at Ra=$2 \times 10^6$ for:  (a) $\gamma=0.19$, (b) $\gamma=0.38$ and (c) $\gamma=0.76$. Note the different scales on the vertical axes.}
  \label{fig:ke_vs_time_M}
 \end{figure}

We now focus on the role of the (nondimensional) condensation parameter $\gamma$, with other parameters fixed. As discussed earlier, changing $\gamma$ modifies the stability of the drizzle solution (as shown in Figure~\ref{fig:drizz_cont}). For fixed $\beta > 1$ and small $\gamma$ the drizzle solution will remain stable whatever the value of $Ra$. As $\gamma$ is increased the gradient of $m(z)$ is altered until is becomes negative and instability occurs if the Rayleigh number is sufficiently high. We thus conduct a series of experiments for moderately high $Ra$ ($Ra = 2 \times 10^6$) and with all other parameters as before, and vary $\gamma$. For $\gamma \lesssim 0.13$ the drizzle solution is linearly stable for all $Ra$. Numerical solutions in fact indicate that for $\gamma \lesssim 0.13$ the solution always returns to the drizzle solution. As $\gamma$ is increased further, steady then unsteady nonlinear solutions are found, and timeseries of the kinetic energy for $\gamma=0.19$, $0.38$ and $0.76$ are shown in Figure~\ref{fig:ke_vs_time_M}. As expected, as $\gamma$ is increased the kinetic energy of solutions increases,  appearing to scale roughly linearly with $\gamma$ for large $\gamma$.
 
Figure~\ref{fig:mean_prof_M} shows the effect of varying $\gamma$ on the mean profiles of buoyancy, relative humidity and $m$. As the heating is increased (by increasing of $\gamma$) the buoyancy of the layer is increased and the relative humidity decreased. The net effect is to increase $m$, with $m$ being well mixed between $z=0.5$ and a thin boundary layer near to the top of the domain. The lower half of the domain remains convectively unstable  (as noted earlier), and whether it becomes neutralized at higher values of the Rayleigh number is a topic for future investigation. 
 
We conclude with a remark on subcritical convection --- i.e., sustained  convection occurring for $Ra<Ra_c$, where $Ra_c$ is the critical Rayleigh number at which the linear instability of the drizzle solution occurs. No such behaviour has been found for this two-dimensional model for an aspect ratio of $20$,  suggesting global (nonlinear) stability for this configuration.
On the other hand, \citet{Pauluis_Schumacher11} and \citet{Weidauer_etal11} did appear to find occurrences of sustained convection in a regime they deemed subcritical by a  different criterion and in a  different system from ours.  \textit{Pace} their results, we are unable to say whether sustained subcritical convection is a general property of moist convection.
 
   \renewcommand*{\putabc}{
   \put(-1.,0.305) {(a)}
   \put(-0.66,0.305) {(b)}
   \put(-0.33,0.305) {(c)}
    }
   \begin{figure}
   \centering
   \includegraphics[width=0.33\textwidth]{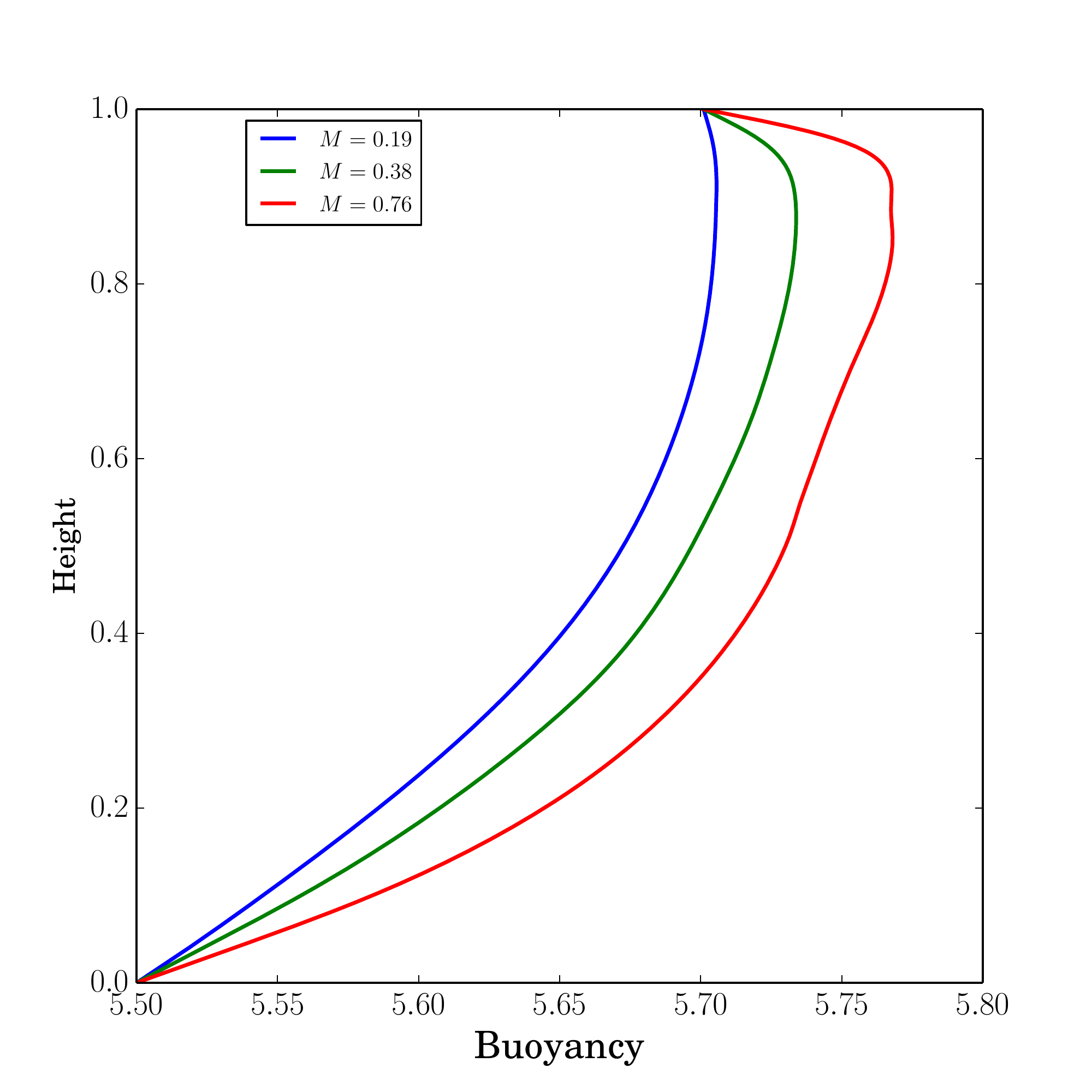}%
   \includegraphics[width=0.33\textwidth]{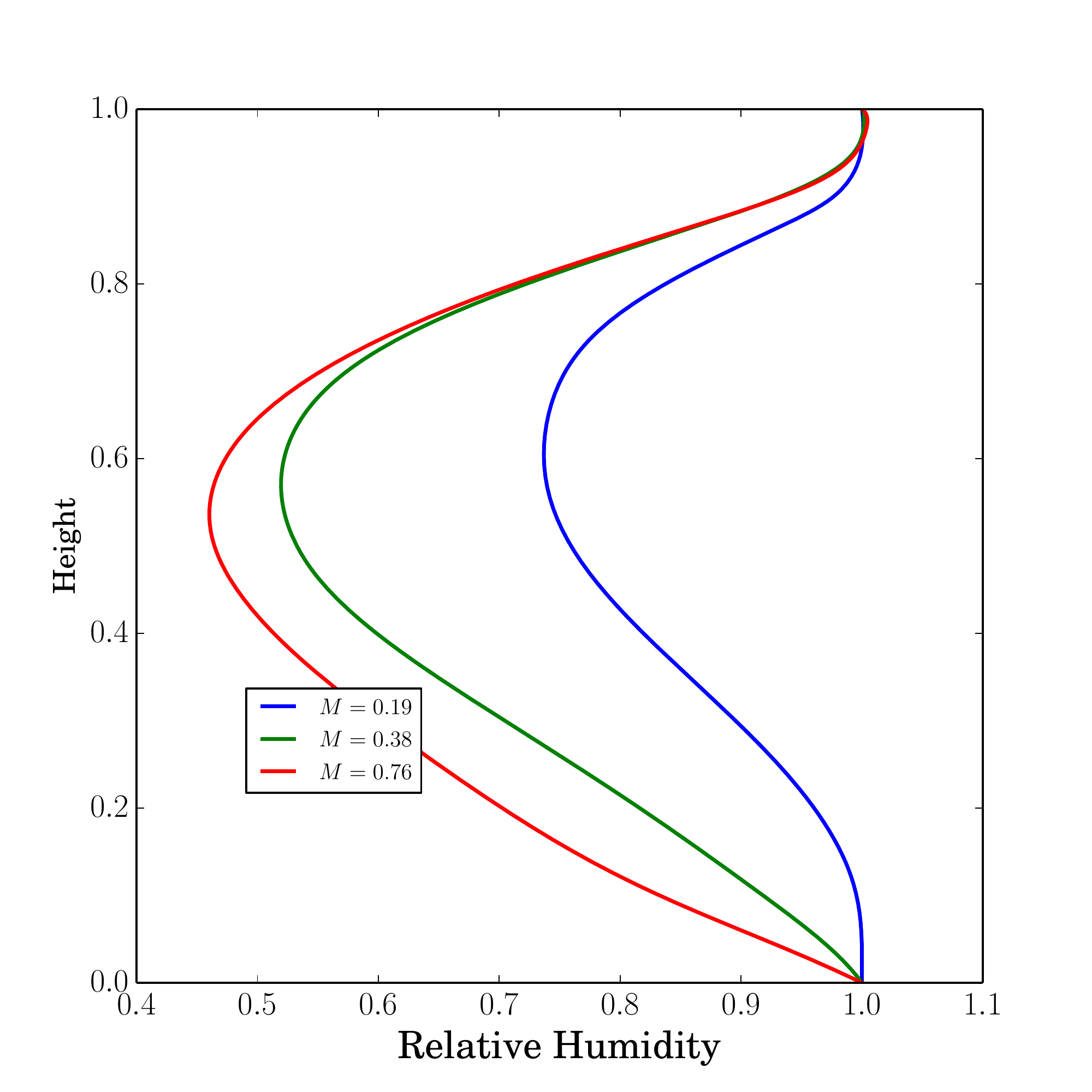}%
   \includegraphics[width=0.33\textwidth]{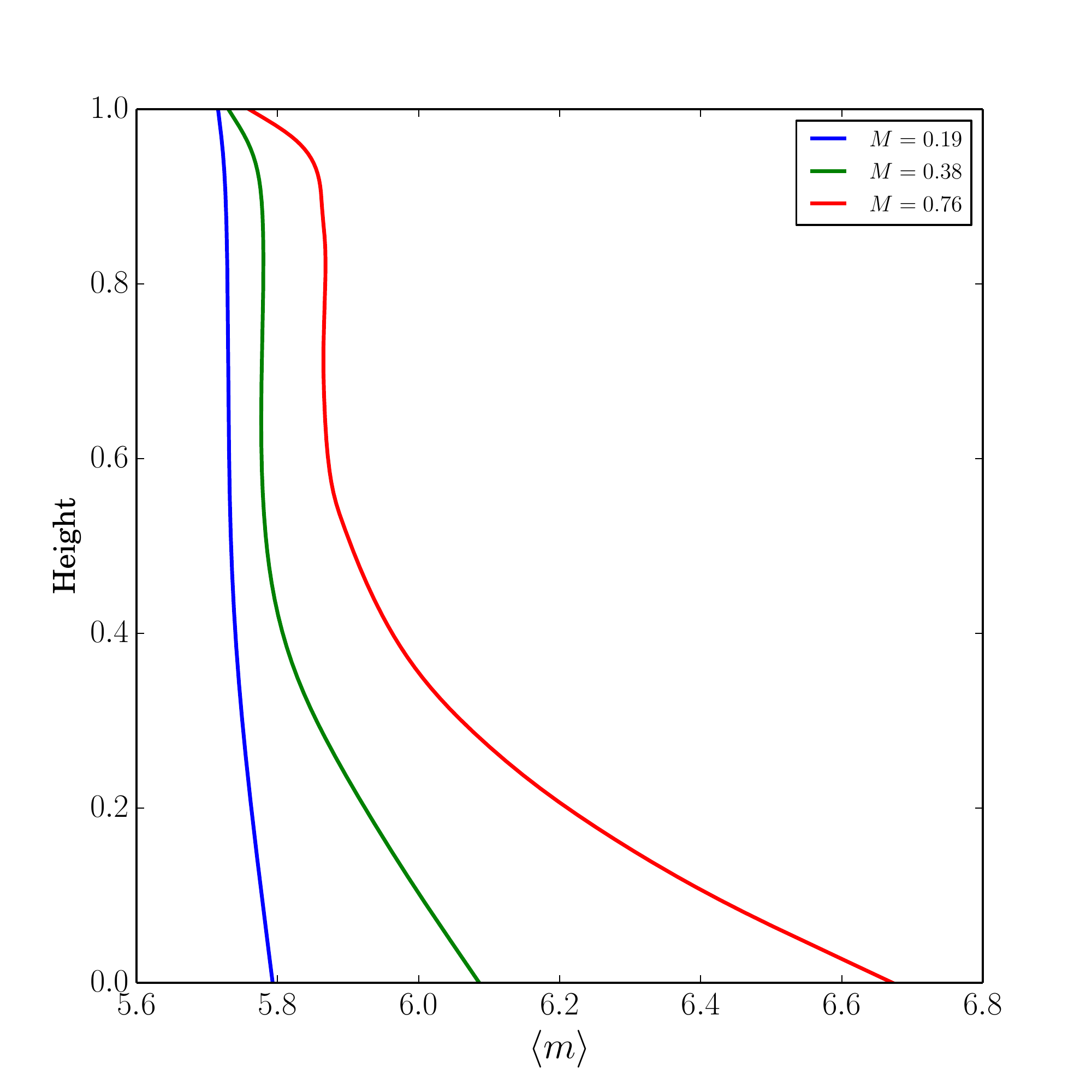}
   \putabc
    \caption{Horizontally averaged profiles for (a) buoyancy, (b) relative humidity and (c) $m(z)$ for $Ra=2 \times 10^6$ and a range of $\gamma$. Note the different horizontal scale for $\langle m\rangle$ from those of earlier figures showing $\avg{m}$.}
    \label{fig:mean_prof_M}
   \end{figure}

\section{Discussion and Conclusions} \label{sec:discussion}
 
In this paper we have presented and begun to analyze a simple model system for moist convection. The literature on moist convection is large but has little overlap with the even larger literature on \RB convection, with a few exceptions as noted in the introduction. Our goals are both to make that connection stronger and to explore moist convection at a fundamental level.   To these ends, we precisely define a relatively simple moist system and look at how that system behaves as the governing nondimensional parameters change, passing from a stationary diffusive-condensing state to a turbulent, time-dependent state. 
 
The system itself, defined by  \eqref{dim.1}, comprises the ideal-gas Boussinesq equations plus an evolution equation for humidity $q$, a definition of temperature and saturation humidity, and a simple recipe for condensation and its effect on buoyancy.   The key nondimensional parameters may be readily identified: in addition to the familiar Rayleigh and Prandtl numbers, the effects of moisture are captured by the parameter  $\gamma$, as in \eqref{ndim.12}, or alternatively a `condensational Rayleigh number', $R_\gamma$, defined in \eqref{Rgamma1}.  Other nondimensional numbers arise from the simplified Clausius--Clapeyron relation and the relation between temperature and buoyancy.   The model system thus simply captures the essential difference between moist and dry convection, reducing to the conventional dry \RB system as the single parameter $\gamma$ goes to zero. 

The  model possesses a drizzle solution of no motion, given by a linear profile of moist static energy $m(z)$.  If the boundaries are saturated then so is the interior, and an exact solution may be obtained in which the diffusion of moisture is balanced by condensation, with the condensation providing a source in the buoyancy equation that is also balanced by diffusion. This balance determines the temperature, and so the saturation value of humidity and so the condensation. The problem is thus inherently nonlinear, although a wholly analytic solution may be written down in terms of a special function.  A drizzle solution may also be found if the lower boundary is not saturated, in which case saturation occurs above some height inside the domain, as in \figref{fig:drizzle-unsat}. The drizzle solution is unstable if the Rayleigh number is high enough and if the values $\gamma$ and the boundary conditions are such that the gradient of $m(z)$ is negative, and a numerically obtained stability boundary is shown in \figref{fig:drizzle-stab}.   For values of parameters that are linearly unstable we always find a solution with motion, and for values of parameters that are linearly stable we always find convergence to the stationary drizzle solution, albeit sometimes after extended transient periods with time-dependent flow. However, we ascribe no generality to this result. 

Numerical solutions of the full system show that, as the Rayleigh number increases, the solution transitions from the steady drizzle solution to a state with a small number of time invariant plumes, and then to a time-dependent and ultimately a turbulent state, with plumes generating gravity waves that remotely trigger other plumes and so on. Typically, solutions are less steady that corresponding solutions at similar Rayleigh numbers in the dry problem, with the plume separation determined by the drying effects of downwelling between the plumes. The plumes are completely saturated when laminar,  but entrain dry air into their flanks when unsteady, more so as the Rayleigh number further increases into a more turbulent regime, with the associated gravity waves giving rise to distinctive patterns of relative humidity (\figref{fig:scatter}).  Overall, the domain becomes dryer in the interior as the system becomes more unstable, although  the simulations may be suggesting that the vertical structure of the relative humidity is tending to a well-defined limit (\figref{fig:mean-rh}).  Clearly, it would be of interest to examine all these issues in three dimensions and at higher Rayleigh number, with the simplicity of the system suggesting that the parameter dependence of the solutions, including such things as the plume width and separation, could, in principle, be unambiguously determined. 

The parameter dependence of the moist system at very high $Ra$ is of particular interest. In the dry problem, and with smooth upper and lower boundaries, theoretical, numerical and laboratory studies all suggest that the solutions always depend upon Rayleigh number; that is, there is no Rayleigh-number independent ultimate regime. It seems unlikely that the presence of moisture alone will alter this conclusion, since the moisture must still enter the domain diffusively through the lower boundary. It is, however, widely believed that the properties of convection in Earth's atmosphere do not, in fact, depend on the molecular diffusivity and viscosity, and properly understanding this issue remains a significant challenge.  We might conjecture that radiative forcing, which provides a source of buoyancy independent of diffusivity, could remove the dependence of atmospheric convection on molecular diffusion and viscosity. Other possibilities exist, such as a rough boundary and turbulence in the boundary layer generated independently of convection, as suggested by success of the Monin-Obukhov theory, which has no explicit dependence on molecular properties but which does rely on a roughness length.

Many other studies need to performed with the model presented here to better understand it.  A well defined pathway toward more realism may also be defined: using an anelastic model instead of a Boussinesq one, adding liquid water and re-evaporation, removing the effects of the upper boundary, and/or adding radiation all lead toward the more complex type of model commonly used in moist atmospheric convection, and we hope that other investigators will also pursue these and related problems. By ensuring that any more complex model is precisely defined and is connected in a direct fashion to a more simple model, one may hope that the results obtained are both reproducible and understandable.

\subsubsection*{Acknowledgements}
Our simulations use the Dedalus framework  (\citealp{Burns_etal16} and http://dedalus-project.org).  We thank the Dedalus team for their fine software and especially Jeff Oishi  for his endless patience. We also thank Jacques Vanneste for drawing our attention to the Lambert W function,  Jan Sieber for conversations and pointing out an error in an earlier draft, and Kerry Emanuel and three anonymous reviewers for their comments. This work was funded by NERC under the Paracon Program via grants NE/N013123/1 and NE/N013840/1 to the Universities of Exeter and Leeds. GKV and DJP also acknowledge support through
Royal Society Wolfson Research Merit Awards, and SMT acknowledges support from the ERC.

\section*{Appendix: Other Nondimensionalizations}

\subsection{Buoyancy-based nondimensionalization}

A buoyancy based nondimensionalization uses the buoyancy differences across the layer rather than  diffusion to scale time, and hence velocity, and 
\begin{equation}
	\label{ndim.15}
	t_s = \bfracsup{H \theta_0}{g \Delta T}{1/2} , \qquad U = W = \left({g \Delta T H  \over \theta_0}\right)^{1/2} . 
\end{equation}
The other scales are the same as with the diffusive nondimensionalization.  The nondimensional momentum equation is
\begin{equation}
\label{mom.2}
     \DDhat \vbhat = -  \del \phihat +  b\kb +  \bfrac{\Pr}{\Ra}^{1/2} \del^2 \vbhat, 
\end{equation}
 and a Reynolds number, $\Rey \sim \Ra^{1/2}$, now appears in the viscous term as expected.  The nondimensional buoyancy equation is
\begin{equation}
	\label{ndim.18}
	\DD \bhat = \gammahat \dfrac{q-q_s}{\tauhat} \H(\qhat - \qhat_s)
		     +   \dfrac{1}{(\Ra \Pr)^{1/2}}\del^2 \bhat , 
\end{equation}
and the nondimensional moisture equation is 
\begin{equation}
	\label{ndim.20}
	\DD \qhat =  \frac{\qhat_s - \qhat}{\tauhat} \H(\qhat - \qhat_s)
	     + \frac{1}{(\Ra \Pr')^{1/2}}  \del^2 \qhat ,
\end{equation}
where $\Pr' = \nu/\kappa_q = \Pr \Sm$. 
The physics equations are unaltered from the case with diffusive scaling.

\subsection{Moisture Based nondimensionalization}
We now use buoyancy created by condensation as a scaling.  Thus we choose
\begin{gather}
    t_s = \bfracsup{H}{q_0 \gamma}{1/2}, \quad U = W = (\gamma q_0 H)^{1/2}, \quad
    B = \gamma q_0, \qtext{so that} b = \gamma q_0 \bhat,
\end{gather}

The momentum equations become 
\begin{equation}
     \DD \vbhat  
    = -  \del \phihat +  \bhat \kb +  \frac{\Pr}{ \Rg^{1/2}} \del^2 \vbhat. 
\end{equation} 
where $\Rg$, the condensational Rayleigh number, is given by
\begin{equation}
\label{Rgamma}
\Rg = \dfrac{\gamma q_0 H^3}{\kappa \nu} 
    = \dfrac{ g L H^3 q_0}{c_p \theta_0 \kappa \nu} .
\end{equation}
Related moist Rayleigh numbers also appear in the analyses of \citet{Bretherton88} and \citet{Pauluis_Schumacher11}. 

The nondimensional buoyancy equation is 
\begin{subequations}
\begin{equation}
	\label{ndim.22b}
	\DDhat \bhat = \dfrac{\qhat - \qhat_s}{\tauhat}  \H(\qhat - \qhat_s)
		     +   \dfrac{1}{(\Rg \Pr)^{1/2}}\del^2 \bhat  ,
\end{equation}
and the nondimensional moisture equation is
\begin{equation}
	\label{ndim.23}
	\DDhat \qhat =  \frac{\qhat_s - \qhat}{\tauhat} \H(\qhat - \qhat_s)
	     + \frac{1}{( \Rg \Pr')^{1/2}}  \del^2 \qhat .
\end{equation}
\end{subequations}
Here $\tauhat$ is the condensation time measured in units of the buoyancy timescale, and this is always a small number in the fast condensation limit.  These equations have no parameter in the heating term in the buoyancy equation due to moisture condensation. However, the effect of the parameter has not been lost for it appears in boundary conditions. For example, if we require a dimensional drop of $\Delta b$ across the domain we implement this with 
\begin{equation}
    \bhat(\zhat=0) = 0, \qquad \bhat(\zhat=1) = {\Delta b \over q_0 \gamma} = {\Delta b c_p \theta_0 \over g L q_0}.
\end{equation}

\bibliographystyle{jfm}
\input rainy.bbl
\end{document}

%% file: mypreamble.tex
\newcommand{\newc}{\newcommand*}

\newcommand{\mycommand}[2]{\providecommand{#1}{} \renewcommand{#1}{#2}}

\providecommand{\bmdefine}{}
\renewcommand{\bmdefine}[2]{\mycommand{#1}{{\bm{#2}}}}
\providecommand{\mathbfup}{\mathbf}

\providecommand{\what}{\widehat}

\newcommand{\medtilde}[1]{\mkern 1.5mu\widetilde{\mkern-1.5mu#1\mkern-1.5mu}\mkern 1.5mu}
\newcommand{\medhat}[1]{\mkern 1.5mu\widehat{\mkern-1.5mu#1\mkern-1.5mu}\mkern 1.5mu}

\providecommand{\bneg}{\mkern -3mu} 
\providecommand{\bnegsup}{\bneg} 

\renewcommand{\;}{\mkern 1.5mu}

\newcommand{\eqnab}%
    {\renewcommand{\theequation}{\arabic{section}.\arabic{equation}a,b}}
\newcommand{\eqnabc}%
    {\renewcommand{\theequation}{\arabic{section}.\arabic{equation}a,b,c}}
\newcommand{\eqnabcd}%
    {\renewcommand{\theequation}{\arabic{section}.\arabic{equation}a,b,c,d}}
	\newcommand{\eqnc}%
	    {\renewcommand{\theequation}{\arabic{section}.\arabic{equation}c}}

\newcommand{\gobble}[1]{\!}
\AtBeginDocument{}

\AtBeginDocument{\renewcommand{\div}[1][]{\nabla_{\! #1}\, \cdot}}

\newcommand{\qtext}[1]{\quad \text{#1} \quad}

\bmdefine{\omb}{\omega}
\bmdefine{\Omb}{\Omega}

\bmdefine{\taub}{\tau}
\bmdefine{\deltab}{\delta}

\newcommand{\er}{\mspace{1mu} \mathrm{e}\mspace{1mu}}

\newcommand{\dg}{\text{\textdegree}\xspace}
\newcommand{\dgc}{\text{\textdegree$\;$C}\xspace}

\newcommand{\eten}[1]{\ensuremath{\times \text{10}^{#1}}\xspace}

\renewcommand{\Pr}{\ensuremath{Pr}\xspace}
\newcommand{\Sm}{\ensuremath{S_m}\xspace}

\newcommand{\Rg}{\ensuremath{R_\gamma}\xspace}

\newcommand{\ro}{\ensuremath{\rho_0} }

\newcommand{\bfrac}[2]{\left( \frac{#1}{#2} \right) }

\newcommand{\bfracsup}[3]{\left(\frac{#1}{#2}\right)^{\bnegsup #3}}

\newcommand{\betahat}{\ensuremath{\what \beta}}

\newcommand{\that}{{\what t}}
\newcommand{\That}{{\what T}}

\newcommand{\ubhat}{\what\ub}
\newcommand{\vbhat}{\what\vb}
\newcommand{\xhat}{{\what x}}

\newcommand{\yhat}{{\what y}}
\newcommand{\zhat}{{\what z}}
\newcommand{\uhat}{\what u}
\newcommand{\vhat}{\what v}
\newcommand{\gammahat}{\what\gamma}
\newcommand{\alphahat}{\what\alpha}

\providecommand{\hatw}{\medhat w}
\newcommand{\bhat}{\what b}

\newcommand{\phihat}{\what \phi}

\newcommand{\qhat}{\what q}

\bmdefine{\jbi}{j}

\providecommand{\shalf}{\tfrac 12}

\providecommand{\mhat}{\medhat m}

\providecommand{\wtil}{\medtilde w}

\mycommand{\tildew}{\wtil}

\newcommand{\tauhat}{\what \tau}

\mycommand{\Zbar}{\overline Z}

\def\dV{\, \mathrm{d}V}

\mycommand{\dz}{\, \mathrm{d}z}

\newcommand{\Fb}{\bm{F} }
\bmdefine\Fb{F}

\newc\Ab{\ensuremath{\bm{A}}}
\newc\Cb{\ensuremath{\bm{C}}}
\newc\Jb{\ensuremath{\bm{J}}}
\newc\Bb{\ensuremath{\bm{B}}}
\newc\xb{{\bm{x}}}
\bmdefine{\pb}{p}
\bmdefine{\qb}{q}
\bmdefine{\ab}{a}
\bmdefine{\Xb}{X}
\bmdefine{\Yb}{Y}

\bmdefine{\sb}{s}
\bmdefine{\Db}{D}

\newcommand{\kb}{\mathbfup{k}}
\bmdefine{\kbi}{k}
\bmdefine{\jbi}{j}
\bmdefine{\Vb}{V}
\bmdefine{\Wb}{W}

\def\vb{{\ensuremath{\bm {v}}}\xspace}
\def\ub{{\ensuremath{\bm {u}}}\xspace}

\bmdefine{\Ub}{U}
\bmdefine{\cb}{c}
\bmdefine{\psib}{\psi}
\bmdefine{\phib}{\phi}

\def\DD#1{{\D#1 \over \D t}}
\newcommand{\DDhat}[1]{{\D#1 \over \D \that }}

\newcommand{\pp}[3][]{{\partial^{#1} #2 \over \partial #3^{#1}}}

\newcommand{\ppp}[3][]{{\partial^{#1} #2 / \partial #3^{#1}}}

\newcommand{\DDD}[1]{\D{#1} / \D t}
\newcommand{\ddd}[3][]{{\d^{#1} #2 / \d #3^{#1}}}

\newcommand{\del}{\nabla}

\newcommand{\cp}{c_{\mkern -1.5mu p}}

\newcommand*{\eps}{\epsilon}

\newcommand{\dd}[3][]{\frac{\d^{#1}#2}{\d #3^{#1}}}
\renewcommand{\d}{\mathrm{d}}

\mycommand{\D}{\mathrm{D}}